# Kinetic effects on the phase behavior and microstructural transitions of a thermoresponsive polymer solution

Pritha Acharya,[a] Riya Karmakar[a] and Khushboo Suman*

Department of Chemical Engineering, Indian Institute of Technology Madras, Chennai, Tamil Nadu, 600036, India

[a] Both authors contributed equally to this manuscript.

* Corresponding Author, email: ksuman@iitm.ac.in

## Abstract

The thermoresponsive behavior of Pluronic® F127 (PF127) triblock copolymer solutions is fundamentally governed by temperature-dependent micellization and complex self-assembly of these micelles. This study systematically investigates the effect of thermal stimuli on the kinetics of phase transition of Pluronic systems during heating and cooling cycles. We employ Differential Scanning Calorimetry (DSC) measurements to investigate the dependence of the micellization temperature on thermal stimuli, revealing that both the micellization temperature and the peak intensity vary systematically with the applied thermal ramp rate. Furthermore, we employ rheological characterization which reveals a sharp sol to soft-solid transition upon heating. Interestingly, we observe a novel multi-step transition during the cooling phase, indicating a more complex reorganization pathway with intermediate metastable states than typically assumed for reversible micellization. We also investigate the effect of thermal cycling on the reversibility and thermal stability of the polymer system. Our findings indicate that the characteristic multi-step cooling transition is transient, gradually weakening with successive thermal cycles. We also present a comprehensive mathematical model which accurately captures the kinetics and multiple step transition in viscoelastic parameters. Significantly, the distinct peaks in Small-Angle X-ray Scattering (SAXS) measurements clearly reveal the evolution from a disordered unimers/micelles state at low temperatures to a highly ordered lattice with long-range spatial correlation at elevated temperatures. We also present a comprehensive phase diagram highlighting the critical role of thermal stimuli and pathways in defining the phase behavior of Pluronic system. This work, therefore, offers essential experimental and theoretical insights into the thermally driven self-assembly, transition kinetics, and microstructural evolution of thermoreversible Pluronic solution.

## Introduction

Over the past few decades, soft materials that sense and adapt to their external stimuli such as temperature, pH, and ionic strength, by reversibly configuring their microstructure, and viscoelastic properties have been widely recognized all over the world.(1-4) This broad family of soft materials now occupy a central role in both fundamental soft matter research and industrial process design and is collectively termed as stimuli-responsive soft materials.(5-7) The responsiveness of these materials towards stimuli arises from the cooperative and relatively weaker intermolecular interactions, including hydrogen-bonding, hydrophobic aggregation, van der Waals attractions, and screened electrostatic interactions that govern their microstructure across nanometre to micron length scales.(1, 2) Therefore, even modest external perturbations can induce substantial and finely tuneable changes in viscoelasticity, phase transition behavior, and microstructure, enabling stimuli-responsive materials to exhibit functionalities beyond those of conventional engineering materials. (7, 8)

Among the diverse stimuli utilised to engineer responsive behavior, temperature stands out to be especially attractive due to its industrial relevance, as it enables continuous and non-invasive tuning of material properties without altering chemical composition. (6, 8, 9) Temperature driven phase transitions in soft materials such as sol-to-gel transitions, order-disorder transformations, and thermally-induced aggregation, are now well established as central to a broad range of industrial processes and product formulations. (8, 9) Recent studies have demonstrated that the thermo-responsive systems show reversible transition between sol-like and solid-like states through temperature modulation alone, offering exceptional flexibility in process design across applications such as pharmaceuticals, food, personal care, enhanced oil recovery, advanced soft robotic and coating technologies. (5, 10, 11) Among thermoresponsive polymer systems, amphiphilic block copolymers are of particular interest due to their ability to undergo temperature-driven self-assembly in selective solvents, leading to unique phase behavior and tuneable material properties. (3, 11, 12) Pluronic triblock co-polymers, also known as poloxamers, consist of poly(ethylene oxide)–poly(propylene oxide)–poly(ethylene oxide) (PEO–PPO–PEO) blocks and exhibit pronounced temperature-dependent self-assembly in aqueous environments. (13) Among this family, Pluronic® F127 (Poloxamer 407), with a molecular architecture $PEO_{100}$–$PPO_{65}$–$PEO_{100}$, has been extensively studied over several decades and remains one of the most widely used thermoresponsive polymers owing to its biocompatibility, non-toxicity, and reversible gelation near physiological temperatures. (12, 14, 15) At low temperatures, Pluronic® F127 (PF127) exists as molecularly dispersed unimers in water, whereas increasing temperature reduces the hydration of the PPO block, driving spontaneous micellization with a hydrophobic PPO core and a hydrated PEO corona. (11) At sufficiently high polymer concentrations and temperatures, the resulting increase in micelle number density and effective volume fraction leads to crowding and intermicellar interactions, giving rise to solid-like mechanical behavior and thermoreversible gelation. (12, 16, 17) These combined features have established PF127 as a model thermoresponsive material and a widely employed platform for injectable hydrogels, transdermal and ophthalmic drug delivery systems, bioadhesive formulations, and rheology modification in pharmaceutical applications. (10, 18)

The temperature-induced phase behavior of aqueous PF127 has been investigated for over three decades using a wide range of experimental techniques, including Differential Scanning Calorimetry (DSC), (19) rheology, (12, 14, 19) Nuclear Magnetic Resonance (NMR), (20) Small-Angle Neutron and X-ray Scattering (SANS/SAXS) (21), and Dynamic Light Scattering (DLS) (22). Early calorimetric studies established that the sol–gel transition is a physical process driven by temperature-induced micellization and packing of spherical micelles, with the PPO blocks playing a dominant role and no involvement of chemical crosslinking. (13, 19) Previous SAXS studies have reported pure hexagonal close packing micellar ordering in lyotropic non-ionic surfactant systems (23), as well as reversible transitions between hexagonal close packing and body centered cubic phases in block copolymers of single component. (24, 25) Subsequent light scattering and SANS studies suggested that the soft solid-state forms through repulsive interactions among micelles, resembling hard-sphere crystallization, accompanied by increasing aggregation number and long-range order despite nearly constant micellar size. (21) Rheological studies previously have shown that PF127 solutions undergo temperature-driven sol–gel and gel–sol transitions whose characteristic temperatures and temperature ranges depend on polymer concentration and thermal protocol, reflecting slow micellar rearrangements and long relaxation times near the transition regime. (26, 27) More recent high-resolution rheological and calorimetric studies have further challenged the notion of a single sol–gel transition, revealing multiple arrested states particularly at higher temperatures including hard gels, viscoplastic soft-solids, and glass-like states. (12, 27, 28) In particular, Suman and Sagar showed that concentrated PF127 solutions undergo a sequence of sol–gel–glass transitions upon heating, with a percolated gel network evolving into a dynamically arrested, glass-like state as micellar volume fraction and intermicellar interactions increase.(12) Collectively, these studies demonstrate that the liquid–soft-solid transition in aqueous PF127 is highly sensitive to concentration, temperature, and experimental protocol, and that the soft-solid state spans a range of distinct arrested microstructural states rather than a single, universally defined gel.(12, 29)

Despite the substantial amount of research devoted to the temperature-responsive phase behavior of PF127, a clear imbalance still remains in the literature. Most studies primarily address heating-induced sol–gel transitions, whereas the reverse process involving cooling and repeated thermal cycling has received comparatively limited attention. (16, 17, 29) Furthermore, the kinetics of the sol-gel and gel-sol transition, in particular is unexplored in the literature. The earliest evidence of such thermal history–dependent asymmetry was reported by Kovacs in his seminal studies on amorphous polymers, where non-monotonic relaxation and memory effects were observed following temperature changes, establishing that heating and cooling pathways need not be equivalent. (30) Later studies on other thermoresponsive soft matter systems, such as particularly dense PNIPAM-based microgel suspensions, have demonstrated strong asymmetry between heating and cooling responses, with cooling often associated with enhanced energy dissipation arising from delayed microstructural reorganization. (31) These observations indicate that once a system enters a mechanically arrested or glass-like state, its return towards equilibrium during cooling can proceed through collective rearrangements spanning multiple timescales, potentially influenced by differences in the available free volume along cooling and heating pathways. (30, 31) Furthermore, even in related rheological studies on modified PF127 systems, heating and cooling cycles have been reported to exhibit distinct mechanical responses; however, the origin of this asymmetry and its implications for microstructural stability during cooling were not explicitly discussed. (32) For pure PF127 and composite systems, although differences between heating and cooling transition temperatures and cycles have occasionally been reported (17, 32, 33), a systematic understanding of how thermal ramp rates and thermal cycling influence the stability and reproducibility of phase transitions, as well as the underlying microstructural evolution, remains unexplored. This gap is particularly relevant for biomedical, pharmaceutical and bioprinting applications, where Pluronic based formulations routinely experience multiple thermal cycles during processing, storage, and where consistent mechanical performance is critical.

In this work, we initially conduct DSC to determine the micellization temperature of PF127 at various thermal ramp rates. Furthermore, we systematically investigate how variation in thermal ramp rate, the number of repeated thermal cycles, and the storage duration since sample preparation influence the phase transition behavior and viscoelastic response of aqueous PF127 dispersions via rheology. (34) Rheological measurements are complemented by SAXS, which reveals the associated structural ordering in PF127 is inherently kinetic in nature. Furthermore, we also present a mathematical model to describe the thermoresponsive viscoelastic behavior of PF127, and it is proposed that this model can be extended to other thermoresponsive soft material systems. Taken together, these results highlight the importance of accounting kinetic effects for identifying transition temperatures and interpreting phase behavior in thermoresponsive soft materials, which offers a practical guidance for the design and industrial processing of such systems.

## Materials and methods

Pluronic® F127 (Poloxamer 407) is obtained from Sigma-Aldrich and used without any further purification. The polymer has a molecular weight of 12,600 g/mol and exhibits a critical micellization concentration (CMC) of about 1000 ppm at room temperature, as reported by supplier. The sample preparation was performed following the cold technique, originally introduced by Schmolka, (35) since PF127 shows enhanced solubility in cold water due to favourable hydrogen bond formation. In this procedure, the required mass of PF127 is weighed and gradually added to Millipore water (resistivity 18.2 MΩ.cm) maintained at 4 °C. The resulting solution with a concentration of 17 wt. % of PF127 is refrigerated at 4 °C for seven days to ensure complete homogenization.

We perform Differential Scanning Calorimetry (DSC) measurements using TA DSC25 – Discovery series instrument to investigate the micellization process of PF127. The measurements were performed over a temperature range from 4 °C to 50 °C. A total of three thermal ramp rates ($k$ = 1, 3, and 5 °C/min) were investigated under both heating and cooling conditions. After completion of each cycle (heating followed by cooling) a rest time of 10 minutes at 4 °C was provided to ensure homogenization of the sample between the runs.

Further to understand the gelation mechanism of PF127, rheological measurements are conducted using an Anton Paar MCR 302e rheometer with serrated concentric-cylinder (CC17) geometry. (34) This geometry comprises of 16.66 mm inner bob, and 18.077 mm outer cup with 0.71 mm gap, along with serrated surfaces designed to minimise wall slip and ensure accurate shear transmission during oscillatory experiments. The temperature control is achieved using a Peltier system integrated with the MCR 302e rheometer, allowing precise thermal regulation over the explored range of temperature in this work 4 °C to 50 °C with a temperature resolution of ± 0.1 °C. The sample was always pre-equilibrated for all the experiments at low temperature (4 °C), where it is in sol state. In this work, oscillatory tests were conducted using a constant stress amplitude of 0.1 Pa, which was confirmed to lie within the linear viscoelastic regime across all temperatures investigated. (12) We perform temperature ramp experiments under oscillatory shear at a constant angular frequency ($\omega$) of 1 rad/s. The temperature ramp rates ($k$) ranging from 0.1 to 5 °C/min were employed for the heating cycles. During the cooling cycle, $k$ from ▯▯0.1 °C/min to ▯▯3 °C/min was explored and faster cooling ramp rates could not be performed due to instrument limitations. We also performed frequency sweep experiments under isothermal conditions at various temperatures. The range of $\omega$ is varied from 0.5 to 50 rad/s while the strain amplitude is maintained at 0.5 %. Moreover, the rheological experiments are performed on day 11, 18, 25, and 33 after the date of preparation of 17 wt.% PF127 sample. Therefore, all the results discussed are based on observations recorded on day 25, unless stated otherwise.

To understand the microstructural evolution of the system, we perform Small-Angle X-ray Scattering (SAXS) measurements using a Xenocs XEUSS 3.0 instrument. The experiments were performed over a temperature range from 4 °C to 50 °C using a Linkam HFSX – 350 – CAP sample environment. Most of the measurements were carried out at an exposure time of 300 s to ensure an optimal signal-to-noise ratio, while selected temperatures of particular interest (20 °C to 25 °C) were probed with a longer exposure time of 600 s to enhance data quality. The sample to detector distance was fixed at 400 mm to access the lower $q$ (scattering vector) range sensitive to larger length scale structures, while the X-ray beam energy was set to 50 keV to ensure reliable scattering intensity across the explored conditions. Additionally, the background correction was performed to ensure the accuracy of the measured scattering profiles.

## Results and discussions

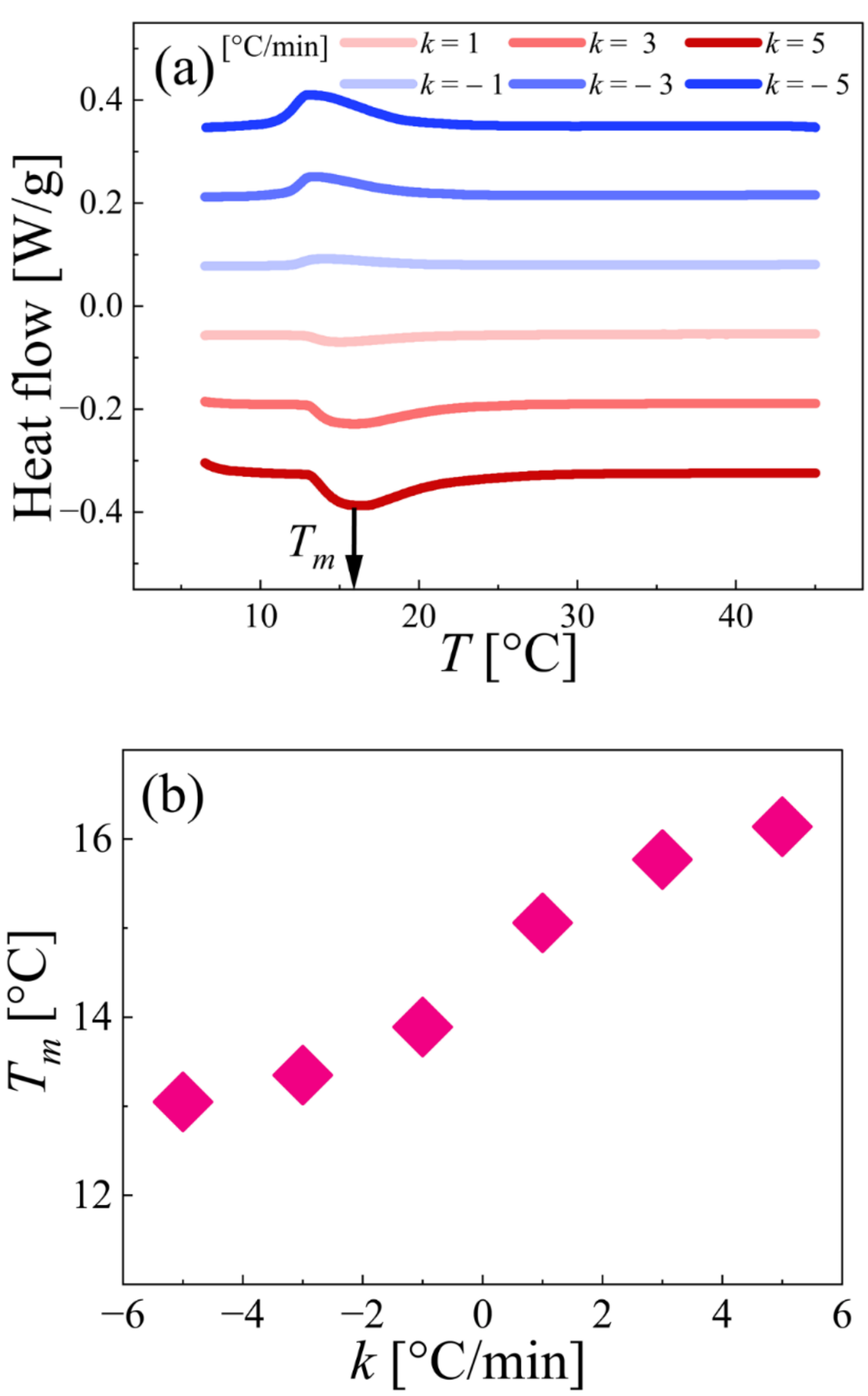

**Figure 1. (a)** The variation in heat flow is plotted as a function of temperature ($T$) with varied ramp rates ($k$) for a 17 wt.% PF127 dispersion measured using a Differential Scanning Calorimeter (DSC) during increasing (shades of red) and decreasing (shades of blue) temperature cycles. The black arrow indicates the micellization temperature ($T_m$). **(b)** The variation in $T_m$ as a function of ramp rates ($k$) is plotted.

In this study, we explore 17 wt. % PF127, a well-established thermoresponsive system with strong relevance to industrial processing and end use applications, where precise temperature control governs performance. When heating the PF127 solution, the initial stage is characterized by micellization. This process involves individual polymer chains, known as unimers, clustering together to form organized structures called micelles. The micellization process is an endothermic process and can be accurately identified using Differential Scanning Calorimetry (DSC) technique. Small amount of the PF127 sample was loaded on the aluminium pans and thermal scans between 4 °C to 50 °C with varying *k* were performed while heating and cooling the sample. It is important to note that a rest time of 5 minutes at 4 °C was provided to ensure homogenization of the sample between the runs. The results of the thermal scans are shown in Figure 1 a, where the variation in heat flow is plotted against temperature during the heating and cooling cycle. The heat flow scan for a heating cycle followed by the cooling cycle confirms the thermoreversible response. It can be observed that the heat flow exhibits a thermal peak corresponding to the temperature at which unimers begin to assemble at all values of the explored thermal ramp rates. The temperature associated with this peak corresponds to the micellization temperature ($T_m$) of the polymeric solution. The micellization temperature is plotted as function of ramp rate in Figure 1 b. It can be observed that during the heating cycle, the value of $T_m$ shifts to higher temperatures. Furthermore, the peak intensity also increases with *k*. This is due to the fact that at slower *k*, the PPO blocks of the polymer chains begin dehydrating along with associating gradually into a micelle and therefore, micellization appears at lower temperature with a smaller peak height. However, at faster *k*, the dehydration in PPO blocks occur together and many unimers arranges into a micelle resulting in a greater peak intensity. A similar kinetic argument applies to explain the behavior observed during the cooling cycle.

The enthalpy associated with the micelle formation is estimated as 3.5 J/g from the DSC experiment at *k* = 3 °C/min. Interestingly, the theoretical value of enthalpy calculated from the reported linear relation $\Delta H = a * (C_{\mathrm{F127}}) + m$, where $a$ = 0.193 J/g, $m$ = 0.43 J/g and $C$ (wt.%) is the concentration of the sample is found to be 3.7 J/g which is in close agreement with the experimental value and previous reported value in the literature. (36) (37) The change in $T_m$ and peak height with thermal *k* suggests that micellization is kinetically governed and not an equilibrium process. The existing studies in the literature have performed DSC to characterize $T_m$ and gelation temperature (for polymer concentration above 20 wt.%) and primarily focused on the dependence of $T_m$ on polymer concentration. (33) From what we can gather, we believe that such thermal ramp-rate-dependent variations in both the peak intensity and $T_m$ in Pluronic system is not yet reported. Although DSC analysis provides valuable insight into the micellization process, it does not provide any evidence regarding the phase transition of PF127. Therefore, comprehensive rheological measurements are conducted to characterize the sol to soft-solid transition of the system.

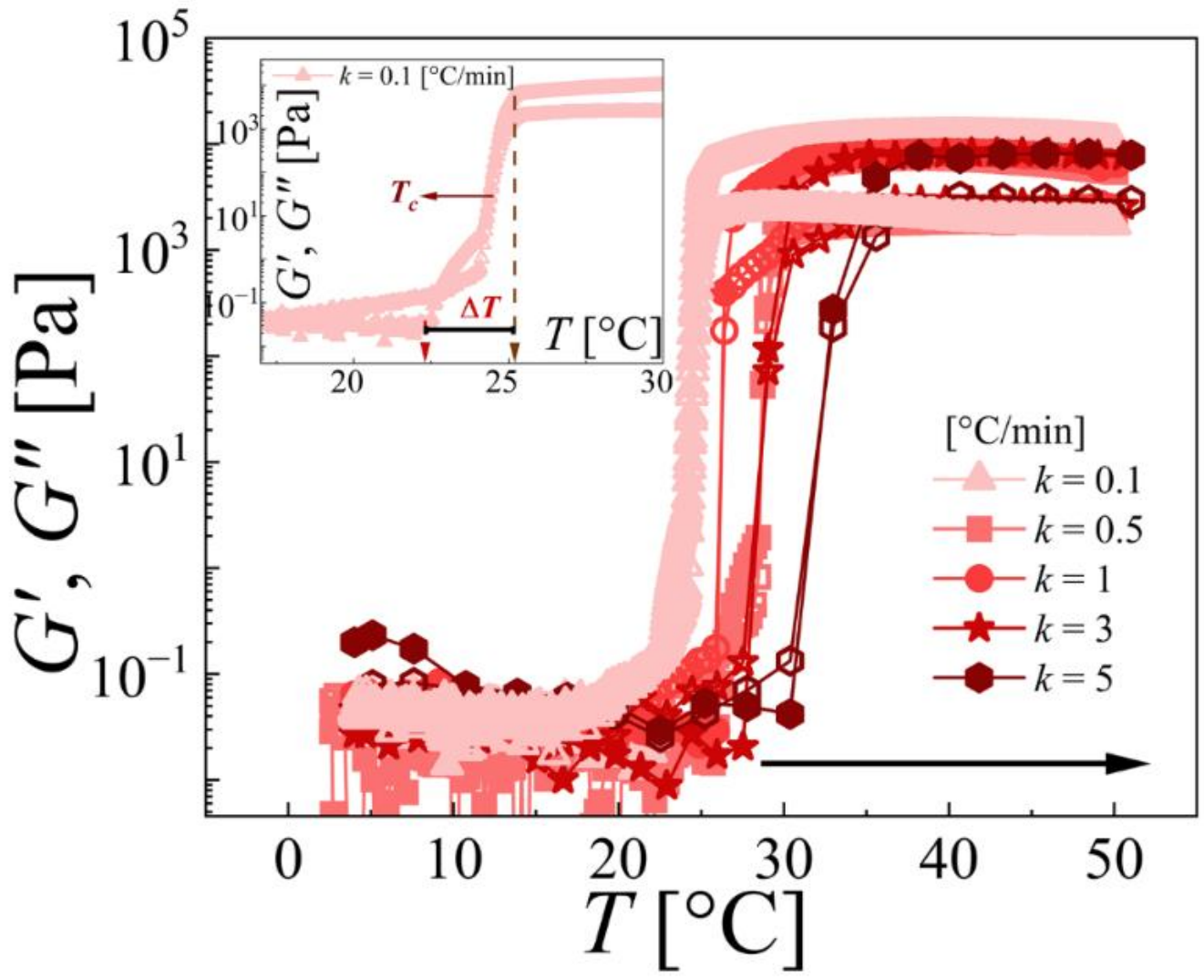

**Figure 2.** The variation in elastic modulus (*G′*, closed symbols) and viscous modulus (*G″*, open symbols) of 17 wt. % PF127 is plotted as a function of temperature (*T*) during heating cycle at varying temperature ramp rates (*k*). Each symbol represents *G′* and *G″* at different *k*, highlighting the influence of heating rate on the viscoelastic behavior of the system. The solid lines are included only to guide the eye and the black solid arrow indicates the direction of temperature followed during a heating cycle. The inset illustrates the methodology used to define the crossover temperature ($T_c$) and the total transition temperature window ($\Delta T$) for heating cycle.

We initiate our rheology study with a focused temperature sweep from 4 °C to 50 °C, to understand the temperature dependent structural evolution of the material system within the operational window of varied processing conditions, as well as storage and transportation environments. In order to understand the changes in the bulk properties of the PF127 dispersion, we subject the sample to heating cycle in a rheometer couette shear cell. In Figure 2, we plot the variation of elastic ($G'$) and viscous ($G''$) moduli as a function of temperature corresponding to a temperature ramp experiment during heating from 4 °C to 50 °C. We explore five different $k$ namely 0.1, 0.5, 1, 3 and 5 °C/min as represented by different symbols in Figure 2. It can be observed that at all heating rates, the value of $G'$ and $G''$ at low temperatures is quite low, consistent with the solution remaining in a sol-like state. As temperature increases, $G'$ and $G''$ rise sharply, signalling the formation of a connected micellar network (sol to soft-solid transition) and it then crosses over leading to a final soft-solid like response at higher temperatures. Interestingly, during the heating cycles, we also observe that the structural rearrangements accompanying the transition of the system from a sol to a soft-solid-like state vary systematically with increasing $k$, as shown in Figure 2. We define the crossover temperature ($T_c$) as the temperature where $G'$ and $G''$ are identical, which is represented by a solid arrow in the inset of Figure 2. This point of crossover has been routinely used to characterize the gelation transition point in soft material systems. (26) Additionally, the inset of Figure 2 illustrates the methodology used to determine the transition temperature window ($\Delta T$) for the heating cycle. We define $\Delta T$ as the temperature interval between the point at which $G'$ first deviates from a constant baseline at low temperature and the temperature at which it reaches a plateau at higher temperatures. The estimation of $\Delta T$ helps in comparing the effect of $k$ across all the temperature profiles used in this study. We observe that irrespective of the applied $k$, the system consistently achieves the same plateau modulus at higher temperatures.

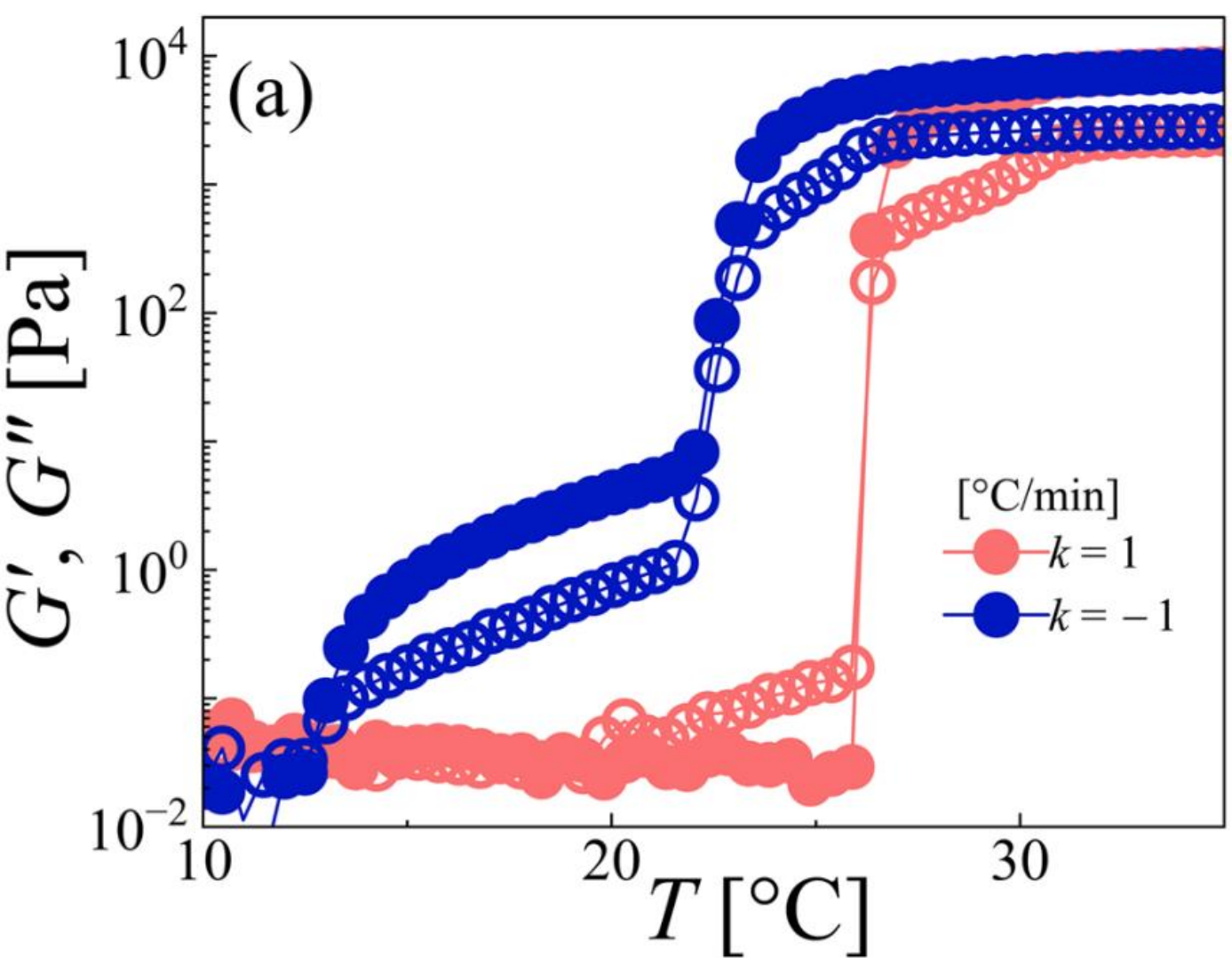

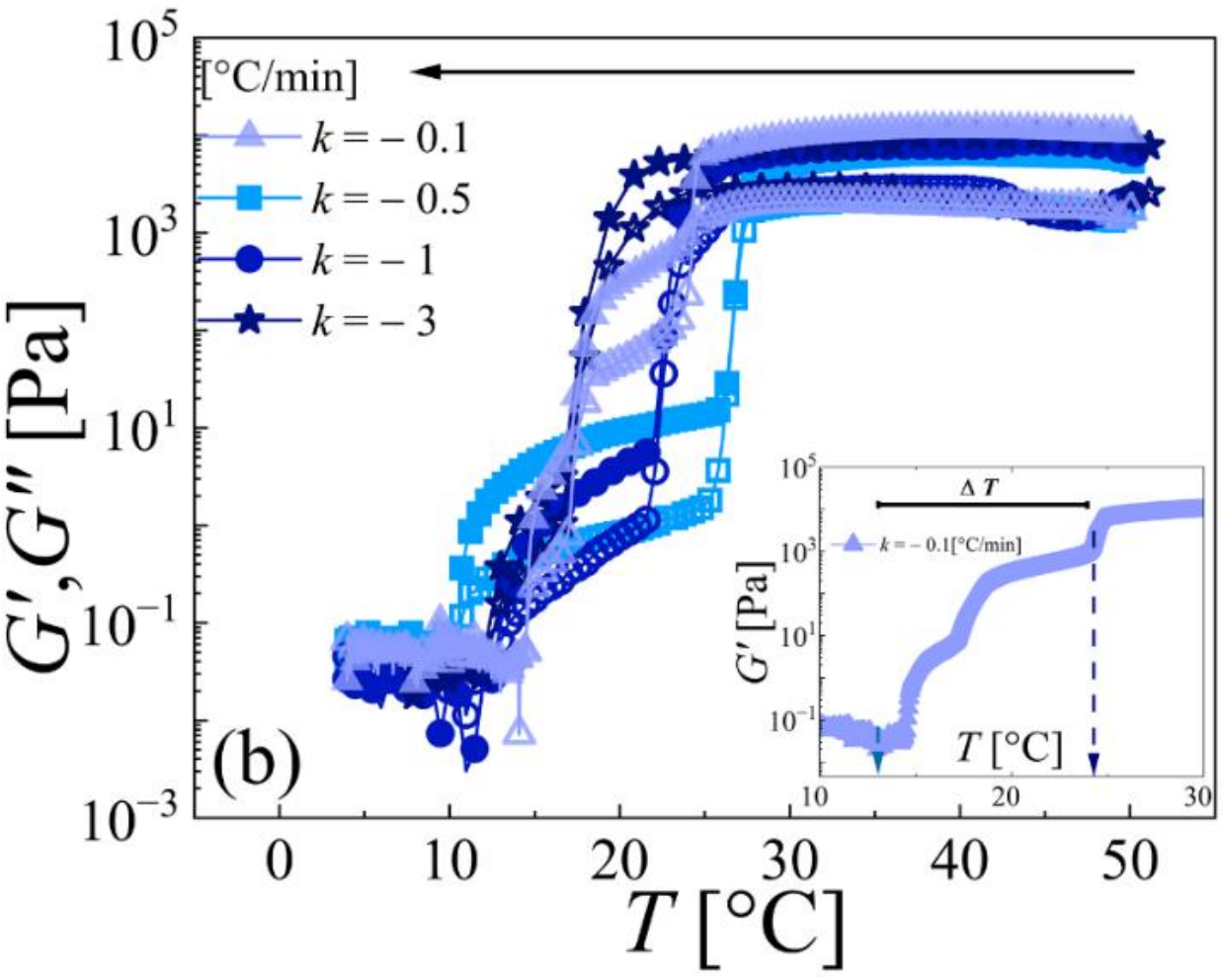

**Figure 3.** The variation in elastic modulus (*G′,* closed symbols) and viscous modulus (*G″,* open symbols) of 17 wt. % PF127 dispersion is plotted as a function of temperature (*T*) during **(a)** both heating and cooling cycles at a ramp rate (*k*) = $\pm$ 1 °C/min and **(b)** cooling cycle at varying *k*. Each symbol represents *G′* and *G″* at different *k*, highlighting the influence of cooling rate on the viscoelastic behavior of the system. The black solid arrow indicates the direction of temperature followed during a cooling cycle and the solid lines are included only to guide the eye. The inset illustrates the methodology used to define the total transition temperature window ($\Delta T$) for cooling cycle.

While most of the literature studies focus on phase transitions of PF127 during heating cycles, industrial applications typically involve both heating and cooling step, which can significantly influence the viscoelastic response of the system. Therefore, in order to understand the effect of cooling step, the heating cycle is followed by cooling cycle from 50 °C to 4 °C. We monitor the evolution of *G′* and *G″* as a function of both increasing and decreasing temperature at *k* = $\pm$ 1 °C/min in Figure 3 a. It is observed that the moduli curves follow distinct pathways during the cooling cycle when compared to their corresponding heating cycle. At the onset of cooling cycle, the system exhibits high moduli, indicative of a soft-solid like state, consistent with the values attained at the end of the heating cycle at 50 °C. As the temperature decreases, both *G′* and *G″* of the cooling cycle progressively decrease, reflecting the gradual weakening of network structure and transition toward a sol-like state. Interestingly, we observe that the decrease in the moduli during a cooling cycle exhibits a novel multiple-step decay as opposed to the near single-step transition observed during the heating cycle. Additionally, we can also observe the presence of a clear hysteresis, as illustrated in Figure 3 a, where the evolution of $G'$ during cooling cycle deviates from that during heating. Although the moduli converge to almost identical values at lower and higher temperatures, the pathways of both the cycles vary remarkably, predominantly within the transition regime at intermediate temperatures. This behavior suggests that the underlying microstructural evolution of the thermoresponsive system of 17 wt. % PF127 does not follow the identical pathway during heating and cooling cycle. The asymmetry between heating and cooling cycle emerges from the different activation barriers for network assembly and disassembly, leading to kinetically distinct pathways. During heating, the network formation requires diffusion, collision, association and percolation of the micelles. However, during the cooling cycle, the network weakening occurs via micelle disintegration and disengagement.

This multiple-step decay in moduli during the cooling cycle has not yet been reported or discussed in the literature for thermoresponsive systems. This motivates us to further investigate the cooling cycle, by exploring different values of $k$ such as − 0.1, − 0.5, − 1, and − 3 °C/min, as represented by different symbols in Figure 3 b. Interestingly in Figure 3 b, the cooling cycles corresponding to different $k$ follow distinct pathways, as evident by the variation of moduli as temperature decreases. This indicates that the structural evolution and weakening of the micellar network of the system are influenced by the cooling rate. Similar to heating cycle, we define the crossover temperature ($T_c$) as the temperature where $G'$ becomes equal to $G''$. However, owing to the presence of multiple transitions during the cooling cycle, we determine distinct $T_c$ associated with each transition where $G' = G''$. The first transition observed during cooling occurs at a relatively higher temperature which is denoted as $T_{c1}$, followed by subsequent transitions identified as $T_{c2}$ and $T_{c3}$ at intermediate and lower temperatures respectively.

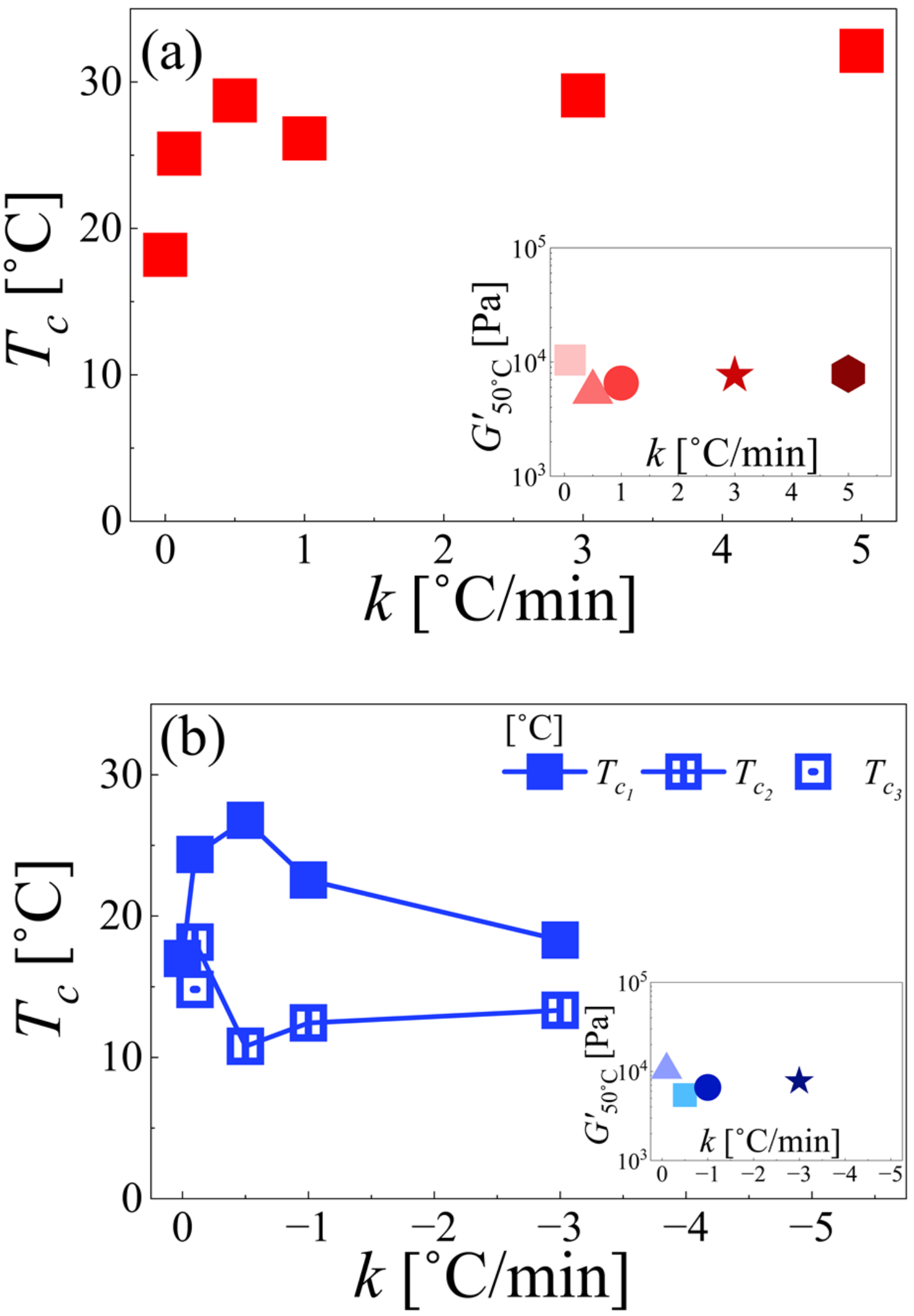

**Figure 4.** The variation in crossover temperature ($T_c$) of 17 wt. % PF127 dispersion is plotted as a function of temperature ramp rate ($k$) with an inset representing $G'_{50°C}$ as a function of $k$ **(a)** during heating cycles and **(b)** during cooling cycles where the solid lines are incorporated to guide the eye. The solid square, square with a vertical line and square with a dot represents $T_{c1}$, $T_{c2}$, and $T_{c3}$ respectively.

In various industrial conditions, materials are frequently exposed to varying thermal ramp conditions, which can influence their structural and viscoelastic response. Motivated by these considerations, along with the distinct behavior observed in Figure 2 and 3, we now systematically investigate the dependence of crossover temperature(s) ($T_c$) on $k$ for PF127 dispersions under both heating and cooling conditions. In Figure 4 a, for the heating cycle, we observe an overall increase in $T_c$ with increasing $k$ (0.1, 0.5, 1, 3, 5 °C/min) characterized by a slight overshoot in $T_c$ at $k$ = 0.5 °C/min. We interpret the rate-dependent shift in gelation as a kinetic effect. The slow thermal ramps provide micelles sufficient time to form, grow and reorganize at each temperature, thereby resulting in a lower gelation temperature as it promotes self-assembly. However, upon increasing the $k$, the re-organization of the micelles gets shifted to an increased temperature value. The faster thermal ramps provide insufficient time for the micelles to thermally equilibrate; consequently, an additional thermal input is required before a percolated micellar network can form. A minor deviation at $k$ = 0.5 °C/min could arise from multiple reasons. On one hand, it could indicate a genuine kinetic effect: at this intermediate $k$, micelles may form only partial or loosely connected structures, and additional time or thermal energy is required for these assemblies to organize into a fully percolated network. This kinetic hindrance can temporarily delay the macroscopic structural transition, leading to the observed increase in $T_c$. On the other hand, it may simply reflect normal experimental scatter, which does not significantly affect the overall trend in $T_c$ with heating rate. Such behavior highlights the sensitivity of micellar gel systems to thermal ramp rates and suggests that intermediate heating rates may induce transient or metastable network states that are not observed at very slow or very fast rates. For a more detailed understanding of this phenomenon, comprehensive scattering experiments (such as small-angle X-ray or neutron scattering) under continuous thermal change would be valuable in the future. Importantly, the final high-temperature (50 °C) plateau modulus, shown in the inset of Figure 4, remain comparable across all $k$, suggesting that the final network stiffness is insensitive to the heating rates. The overall shift in $T_c$ with $k$ suggests that $T_c$ which is generally considered as the definitive point of phase transition, is not a fixed value but rather emerges from a delicate interplay between micelle rearrangement kinetics, network formation timescales, and thermal equilibration.

Similarly, in Figure 4 b, we plot all $T_c$ associated with the cooling cycle where multiple-step transitions are evident, in contrast to the single-step transition observed during heating cycle. Here, $T_c$ associated with the first, second and third transitions, are denoted as $T_{c1}$, $T_{c2}$, and $T_{c3}$ respectively. We observe that $T_{c1}$ increases with increasing magnitude of $k$ up to $k$ = − 0.5 °C/min, followed by a slight decline at faster cooling rates. This non-monotonic behavior suggests a rate-dependent shift in the initial structural transition during cooling cycle under non-equilibrium conditions. Furthermore, it is observed that $T_{c2}$ exhibits a decreasing trend with increasing $k$ from approximately 17 °C to 11 °C at $k$ = − 0.5 °C/min, after which it attains an almost constant value. This indicates that beyond a certain rate threshold, the intermediate transition becomes less sensitive to $k$. Interestingly, $T_{c3}$ is observed only at the slowest cooling rate of $k$ = − 0.1 °C/min, and is absent at all higher ramp rates. We attribute this behavior to the requirement of sufficient time for the system to undergo the third transition, which becomes kinetically inaccessible under faster cooling conditions. As a result, fast cooling ramps suppress these intermediate structural transitions, producing a more abrupt structural collapse. However, slower cooling ramp rates promote sequential microstructural transitions, including partial micellar unpacking, weakening of inter-micellar contacts, and eventual dissolution into unimers. In contrast, the faster cooling rates restrict the time available for such rearrangements, leading the transition to occur abruptly. This behavior highlights the critical role of kinetic control in soft-solid to sol transitions. The observed macroscopic viscoelastic response is not purely determined by equilibrium thermodynamics but strongly depends on the timescale of temperature changes relative to the intrinsic micellar rearrangement and network connectivity. In practical terms, this implies that processing conditions, such as cooling rates in industrial applications, can be tuned to either preserve intermediate network structures or achieve rapid gel collapse, depending on the desired material properties. Additionally, we have plotted $G'_{50°C}$ as a function of $k$ (– 0.1, – 0.5, – 1, – 3 °C/min) for the cooling cycle as an inset of Figure 4 b. The values remain nearly constant across $k$, confirming the same bulk properties at higher temperatures.

This rate-dependent shift in $T_c$ is also observed in other thermoresponsive systems during heating cycle. For instance, Louhichi *et al.* found that the crystallization temperature of Pluronic micellar polycrystals increased with the heating rate used during freezing. (38) Similarly, Franco *et al.* reported from DSC measurements that slower heating yields the phase transition of PNIPAM/PAAc microgels at lower temperatures, whereas increasing the thermal rate shifts the volume phase transition to higher temperatures. (39) Both results are consistent with classical time–temperature–transformation concepts from Kurz and Fisher (2002), (40) which predict that rapid temperature change suppresses nucleation and delays ordering. These reports show that in thermoresponsive polymer systems, rapid heating delays cooperative assembly by pumping energy into the system faster than structure can react. A similar dependence on temperature $k$ has been reported by Kawale *et al.* in dense PNIPAM microgel suspensions, (31) where rapid ramps introduce strong thermal shocks that limit the time available for equilibration, so a higher temperature is required for the system to reorganize into a mechanically connected state. This supports our observation that faster heating shifts *G′* and *G″* crossover to higher temperatures and confirms that the shift arises from rate dependent kinetics rather than a change in the underlying thermodynamic transition. To the best of our knowledge, reports on the effects of cooling ramp rates in thermoresponsive microgel systems are limited. Notably, a very recent study by Kawale *et al.*, (31) describe behavior in dense PNIPAM microgel glasses that slow cooling reveals asymmetric pathways driven by sequential microstructural rearrangements, whereas fast cooling suppresses intermediate states and yields a more abrupt collapse.

By bringing together these observations from Figure 4 a and b, we infer that gelation of 17 wt.% PF127 dispersion depends not only on temperature but also on thermal stimuli. The heating rate sets the temperature at which micellar connectivity begins to develop, whereas the cooling rate governs how clearly we resolve intermediate structural states during cooling. Although the final soft-solid like state at 50 °C reaches a comparable modulus for all heating and cooling protocols, the pathways that lead into and out of this state remain history dependent. These insights highlight that the thermal ramps play a central role in shaping the rheological response of thermoresponsive micellar systems.

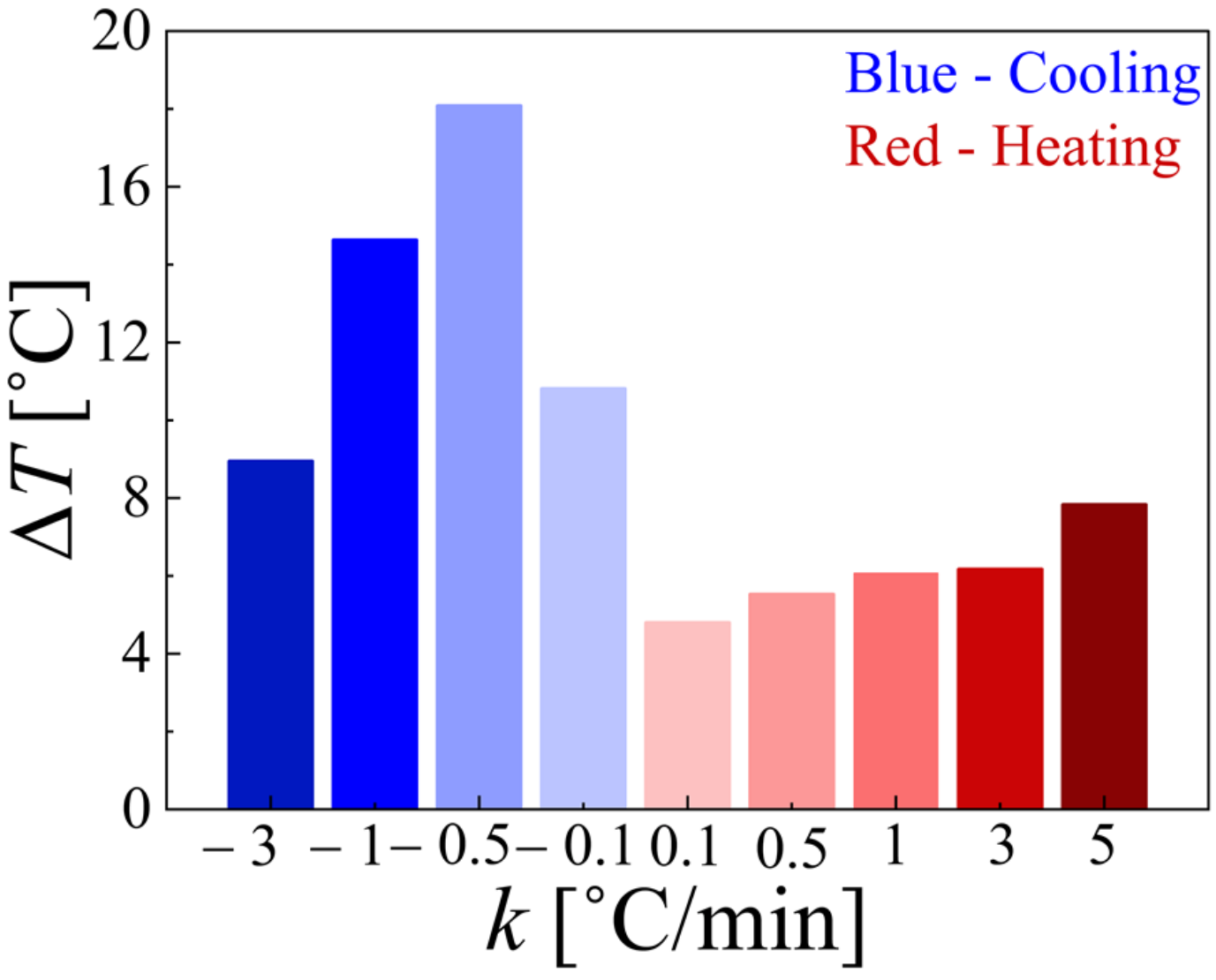

**Figure 5.** The variation of the total transition temperature window ($\Delta T$) for 17 wt. % PF127 dispersion is plotted as a function of temperature ramp rate ($k$) during the heating and cooling cycles.

The observed asymmetry in the evolution of $G'$ and $G''$ during heating and cooling cycles due to the effect of thermal ramps as discussed earlier, motivated us to further quantify the temperature range in the entire transition region where the moduli undergo significant changes. This parameter, transition temperature window $\Delta T$, represents the temperature interval between the point at which the $G'$ deviates from a constant baseline and the temperature at which $G'$ becomes constant again during the heating cycle, and is reiterated here for clarity. Physically, this temperature window represents the range over which the system undergoes structural reorganization and rheological phase transition. As shown in Figure 5, for the heating cycle, $\Delta T$ widens with increasing $k$. At slower $k$, the system experiences a relatively narrow temperature transition window, indicating that the microstructural evolution occurs over a well-defined temperature range. In contrast, at faster $k$, the transition becomes broader, leading to a larger $\Delta T$. We assume this behavior arises because faster heating does not allow sufficient time for the system to reorganize at each temperature. As a result, different structural variations and self-assembly processes, such as micellar rearrangement, growth, and network formation, overlap over a wider temperature range.

For the cooling cycle, the definition of $\Delta T$ remains consistent with that of heating cycle and is defined in the inset of Figure 3 b. As evident in Figure 5, the value of $\Delta T$ exhibits a dependence on the cooling rate. At the slowest cooling rate ($k$ = − 0.1 °C/min), the total $\Delta T$ remains small, likely because slow cooling rate allows the system to relax efficiently at each stage, resulting in sharp and well-defined transitions. At an intermediate cooling rate ($k$ = − 0.5 °C/min), $\Delta T$ shows an anomalous increase, consistent with the behavior observed in our previous rheological analysis. We assume this reflects a competition between the imposed cooling timescale and structural dynamics, leading to broadened and overlapping decay processes. At higher cooling rates ($k$ = − 1 and − 3 °C/min), $\Delta T$ decreases again, possibly due to kinetic suppression of slower timescales. Motivated by the kinetic effect of thermal ramp rate on the viscoelastic moduli, we next develop a mathematical framework to quantitatively describe the observed rheological behavior and its underlying transition characteristics.

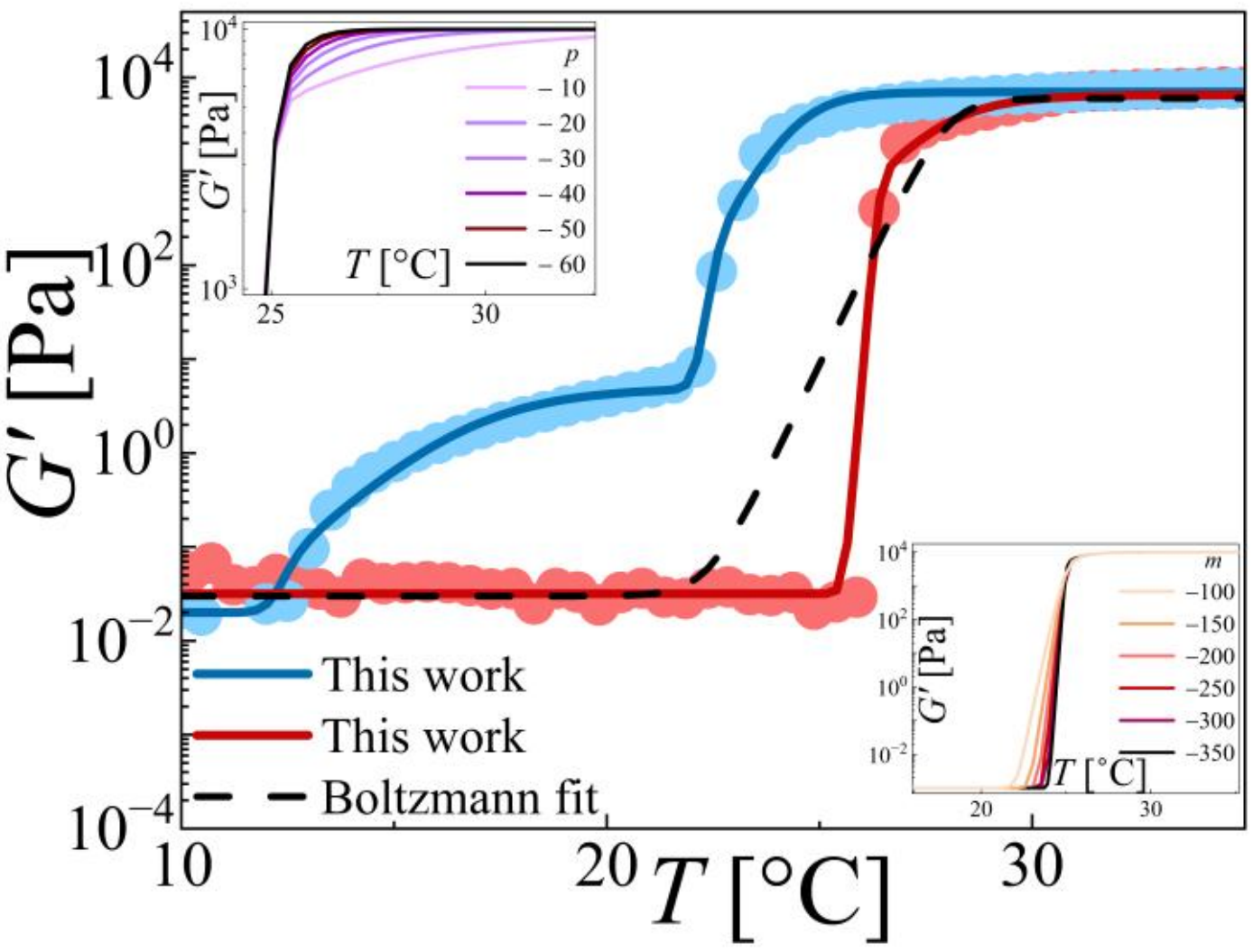

**Figure 6.** The variation of *G′* as a function of temperature during heating and cooling (*k* = ± 1°C/min) cycles of 17 wt. % PF127 dispersion is plotted. The solid red and blue lines represent the fits obtained using the modified kinetic model presented in this work (denoted as this work) for heating and cooling cycles respectively, while the black dashed line represents the Boltzmann sigmoidal fit. The top and bottom inset illustrates the effect of the fitting parameters *p* and *m* on the evolution of *G′* with temperature.

To complement our rheological results, kinetic modelling is employed to provide a quantitative framework for understanding the gelation process captured through rheological experiments. In order to model the rheological behavior of the thermoresponsive polymer system during the heating and cooling steps, we first employ a Boltzmann sigmoidal function, as temperature-induced phase transitions are often described using empirical sigmoidal models to capture S-shaped rheological responses. The Boltzmann sigmoidal curve is given by:

$$G'(T) = G'_0 + \frac{G'_\infty - G'_0}{1+exp\left(\frac{(T-T_C)}{dT}\right)} \ , \tag{1}$$

where $G'$ is the elastic modulus at temperature $T$, while $G'_0$ and $G'_\infty$ denote the asymptotic moduli in the high- and low-temperature limits, respectively which are the known parameters. The parameter $T_c$ is a known parameter and represents the crossover temperature. We fit Eq. 1 to the elastic modulus obtained during the heating cycle in Figure 6. For the sake of simplicity, all subsequent fits presented here are shown only for the elastic modulus; however, fits of comparable quality were also obtained for the viscous modulus. It can be observed that the Boltzmann sigmoidal curve exhibits a much broader and more gradual transition in the region where the system evolves from a liquid-like state to a soft solid-like state, and is therefore unable to adequately capture the experimental data. This limitation arises because Eq. 1 does not contain an adjustable parameter to control the sharpness or breadth of the transition in this regime. We, then take inspiration from the mathematical model proposed by Kopač which builds upon the Ellis model for shear-thinning fluids. (41, 42) In this approach, the temperature dependence of modulus is given by: (42)

$$G'(T) = G'_\infty\left[1+\left(\frac{T}{T_2}\right)^s\right] + \frac{\frac{G'_0}{1+\left(\frac{T}{T_1}\right)^p} - G'_\infty\left[1+\left(\frac{T}{T_2}\right)^s\right]}{1+\left(\frac{T}{T_c}\right)^m}, \tag{2}$$

where, $G'$, $G'_0$, $G'_\infty$ and $T_c$ are known parameter as described previously. The unknown parameters include exponents $p$ and $m$, where $p$ governs the slope evolution at higher temperature plateau, and $m$ controls the sharpness of the transition in $G'$ within the transition temperature range. The influence of $p$ and $m$ on the sigmoidal curve is highlighted in the top and bottom inset of Figure 6 for reference. The characteristic temperatures $T_1$ and $T_2$ describe the evolution of the material response outside the primary transition region, with $T_1$ often associated with hysteresis effects and $T_2$ with high-temperature sensitivity. The exponent $s$ accounts for the change in modulus in the sol-like region.

Taking inspiration from this model, we modify it to better describe our thermoresponsive polymer system. In the present study, the value of modulus at low temperature remains constant and therefore we set the parameter $s = 0$. Under this condition, $G'_\infty$ term becomes temperature-independent, leading to a simplified form of the equation by eliminating two fitting parameters namely, $s$ and $T_2$. Secondly, since we aim to develop a mathematical framework applicable to materials with multiple-step transitions in the viscoelastic moduli, we formulate the model as a summation over the number of individual increase/decrease steps ($i = 1, 2, .., N$). Therefore, the final equation is given by:

$$G' = \sum_{i=1}^{N} G'_{\infty_i} + \frac{\dfrac{G'_{0_i}}{1 + \left(\dfrac{T}{T_i}\right)^{p_i}} - G'_{\infty_i}}{1 + \left(\dfrac{T}{T_{C_i}}\right)^{m_i}}, \tag{3}$$

while retaining the same physical interpretation of all parameters as defined previously and $N$ being the number of steps observed in the transition region. We perform fitting of this simplified model (Eq. 3) using OriginLab and custom-developed MATLAB code. To begin with, we fit Eq. 3 with two fitting parameters (for $N$ = 1) to describe the rheological behavior during heating cycle. It can be observed that simplified form of model fits all the rheological transitions quite well during the heating cycle as shown by red line in Figure 6. The corresponding $p$ and $m$ (fitting parameters) for the heating cycle, for Figure 6 are found to be – 31 and – 300, respectively. Subsequently, for the cooling cycle, we fit Eq. 3 with $N$ = 2 which captures the two-step transition exceedingly well. The corresponding fitting parameters for the cooling cycle, namely $p_{i=1,2}$ and $m_{i=1,2}$, are found to be $p_1 = -12.59$ , $p_2 = -40.72$ and $m_1 = -51.76$ , $m_2 = -145.91$, respectively. We also fit Eq. 3 to all the rheological data obtained at all $k$ during heating and cooling. The corresponding fits and fitted parameters are shown in the Supplementary Information as Figure S 1 to 3. The model fits our rheological data for both heating and cooling cycles, remarkably well (with $R^2$ > 0.9 for all cases). This highlights the ability of our modified Eq. 3 to successfully capture the sharp transition and the inherent hysteresis of 17 wt.% PF127. A key advantage of this model is that the framework can be readily extended to other thermoresponsive materials to capture multiple-step transitions observed in the viscoelastic moduli, where the material undergoes successive structural changes rather than a single transition. Since all measurements are performed under finite $k$, the transitions occur under non-equilibrium conditions; therefore, the true intrinsic transition temperature should, in principle, be defined in the limit $k = 0$, where the system is allowed to follow quasi-equilibrium pathways.

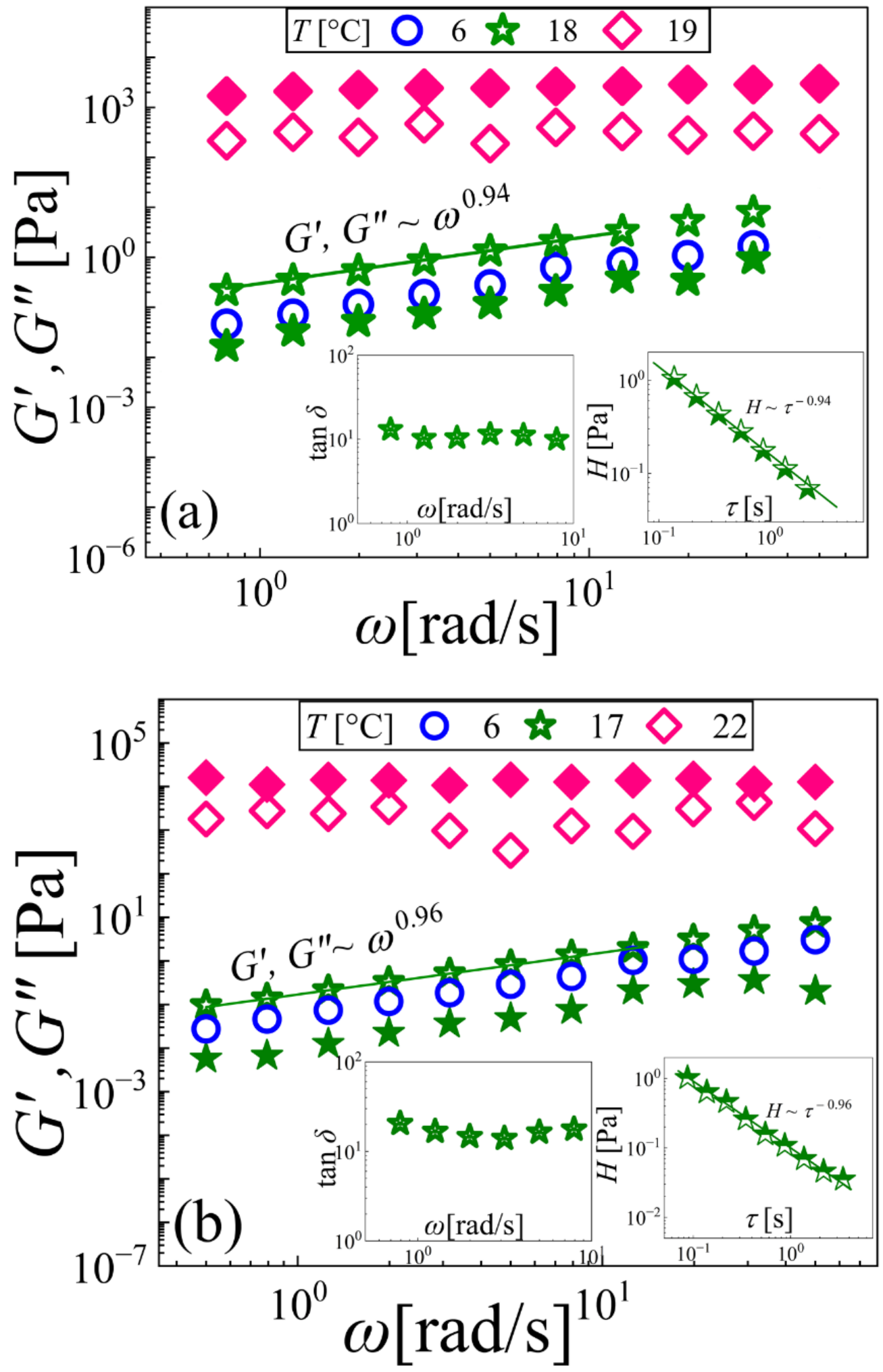

**Figure 7.** The variation of elastic modulus ($G'$, closed symbols) and viscous modulus ($G''$, open symbols) is plotted against angular frequency ($\omega$) for a 17 wt.% PF127 dispersion at **(a)** increasing temperatures and **(b)** decreasing temperatures during the process of gelation. The green straight lines indicate power law scaling behavior in the critical gel state. The left inset shows the variation of loss factor (tan $\delta$) as a function of $\omega$, while the right inset represents the continuous relaxation time spectrum ($H$) as a function relaxation time ($\tau$) at the critical gel temperature ($T_g$) in which the solid line represents the fit given by Eq. 5.

In order to understand how the phase transition takes place under quasi-equilibrium conditions ($k = 0$), where the system is allowed to follow near-equilibrium pathways, we perform isothermal experiments. The frequency is varied between 0.5 − 50 rad/s while keeping the temperature constant and after the completion of one frequency sweep, the temperature is increased to the next target temperature where another frequency sweep in the same window is performed at constant temperature. The isothermal frequency sweep results at different temperatures achieved during the heating protocol on the 17 wt.% PF127 dispersion are shown in Figure 7 a, where the evolution of the $G'$ and $G''$ is plotted as a function of angular frequency ($\omega$). At lower temperature (6 °C), the system displays a sol-like response characterized by the presence of $G''$ which scales approximately linearly with frequency ($G'' \sim \omega$), and $G'$ being below the instrument limit, a characteristic behavior of a viscoelastic fluid in the pre-gel state. As the temperature increases (18 °C), both moduli increase significantly, signalling the onset of structural changes within the system. Notably, $G'$ and $G''$ exhibit identical frequency dependence ($G' \sim G'' \sim \omega^n$), where $n$ represents the critical relaxation exponent that characterizes the power law scaling behavior of the system. (28, 43, 44) The identical scaling of viscoelastic moduli on frequency signifies the emergence of space spanning percolated network of the system. (43-45) Upon further increasing the temperature (19 °C), the system undergoes a clear transition to a soft-solid like material, characterized by $G' > G''$ across the entire frequency window. In this state, the $G'$ approaches frequency independence, demonstrating that the system behaves predominantly as an elastic solid with the establishment of a fully connected network. The left inset of Figure 7 a further highlights the evolution of the loss factor ($\tan\delta = \frac{G''}{G'}$) as a function of angular frequency ($\omega$) at the critical gelation temperature ($T_g$). We observe that at $T_g$ the convergence of tan $\delta$ across all frequencies confirm the frequency-invariant response, thereby reinforcing the identification of the critical gel point.

Furthermore, to connect the observed rheological response to the underlying microstructure of PF127, we compute the continuous relaxation time spectrum $H(\tau)$, from the rheological measurement data plotted in Figure 7 a. The continuous relaxation time spectrum $H(\tau)$ is defined such that $H(\tau)\, \mathrm{dln}\, \tau$ represents the contribution to the modulus arising from relaxation modes whose logarithmic relaxation times lie between ln $\tau$ and ln $\tau + d\ln\tau$. (46) The spectrum $H(\tau)$ plotted in the right inset of Figure 7 a, which can also be derived from $G''$, for which an approximate relationship has been reported by Tschoegl (1989) (47):

$$H(\tau) \approx \frac{2G''}{\pi}\left[1 + \frac{1}{2}\frac{d\ln G''}{d\ln\omega}\right] \text{at } \tau = \frac{1}{\sqrt{3}\omega} \quad \cdots \quad \frac{d\ln H}{d\ln\tau} > 0,$$

$$H(\tau) \approx \frac{2G''}{\pi}\left[1 - \frac{1}{2}\frac{d\ln G''}{d\ln\omega}\right] \text{at } \tau = \sqrt{3}\omega \quad \cdots \quad \frac{d\ln H}{d\ln\tau} < 0. \tag{4}$$

The right inset of Figure 7 a represents the relaxation time spectrum *H* ($\tau$), calculated using Eq. (5), plotted as a function of relaxation time ($\tau$) at the characteristic temperature points identified in Figure 7 a. For a system at the critical gel point, the continuous relaxation time spectrum is described by (8):

$$H(\tau) = \frac{S}{\Gamma(n)}\tau^{-n}, \qquad \tau_0 \leq \tau < \infty \tag{5}$$

where *S* and *n* are material-specific parameters. In the right inset of Figure 7 a, Eq. (5) is superimposed as solid line for the critical gel temperature of 18 °C. We observe that at the critical gel point, the relaxation spectrum exhibits a power-law dependence on relaxation time, $H(\tau) \sim \tau^{-n}$, manifested as a negative slope on a double-logarithmic plot. Additionally, the associated power-law exponent matches the $n$ = 0.94 obtained from the tan $\delta$ analysis during the heating cycle plotted in Figure 7 a.

Figure 7 b represents the results of isothermal experiments obtained at decreasing temperatures during the cooling cycles of the 17 wt.% PF127 dispersion, where $G'$ and $G''$ are plotted as a function of $\omega$. At higher temperature (22 ℃), the dispersion exhibits a soft-solid like state characterized by $G' > G''$ across the measured $\omega$ range. As the temperature decreases (17 ℃), both moduli reduce in magnitude, reflecting a progressive weakening of the gel network. At this intermediate state, the rheological data reveal a power-law dependence on $\omega$, with $G'$ and $G''$ scaling identically ($G' \sim G'' \sim \omega^n$). (43, 44, 48) The scaling behavior signifies the critical gel point during the cooling step, where the system undergoes a structural transition from a viscoelastic gel to sol-like state. Further decreasing the temperature (6 ℃) results in the disappearance of the network connectivity, and the system reverts to a sol-like state. In this regime, $G'' > G'$**,** and $G''$ exhibits an approximately linear frequency dependence ($G'' \sim \omega$), which is consistent with the viscous behavior in the pre-gel state. The left inset in Figure 7 b shows tan $\delta$ as a function of $\omega$ at the critical gel state, where its frequency independence across all $\omega$ confirms the identification of the critical gel point during cooling cycle. Additionally, the continuous relaxation time spectrum in the right inset of Figure 7 b exhibits a power law dependence on relaxation time at the critical gel temperature. The corresponding power law exponent is consistent with the $n$ = 0.96 obtained from tan $\delta$ analysis of Figure 7 b during the cooling cycle. Taken together, we can conclude that all these parameters calculated exhibit identical trends for both heating and cooling cycles.

To ensure clarity, we would like to emphasize that $T_c$ , defined in this study as the temperature at which $G'$ [illegible] $G''$ during temperature sweep measurements for both heating and cooling cycles at finite ramp rates ($k \neq 0$), is distinct from $T_g$ obtained from the Winter–Chambon criterion under isothermal conditions ($k$ = 0) from frequency sweep experiments. In Figure 4 a and b, we plot $T_g$ from isothermal frequency sweep experiments corresponding to $k$ = 0 in order to enable a direct comparison between quasi-equilibrium gelation and rate dependent transitions. It can be observed that $T_g$ (obtained under isothermal conditions) differs significantly from $T_c$ (obtained from continuous temperature ramp experiments) values for the same system under investigation during both heating and cooling protocol. The difference arises primarily because $T_g$ reflects a near-equilibrium transition measured under conditions that allow sufficient time for structural recovery, whereas $T_c$ is inherently influenced by the imposed temperature ramp rate and therefore captures a kinetically shifted transition. As a result, $T_c$ values are sensitive to thermal history and rate-dependent effects, while $T_g$ represents a more intrinsic material response under quasi-equilibrium conditions. This comparison highlights the limitation of using $T_c$ as a material property to define the phase transition, as it is not an intrinsic constant and varies significantly with the applied thermal stimuli. Therefore, $T_g$ provides a more reliable and physically meaningful parameter for capturing the true transition point.

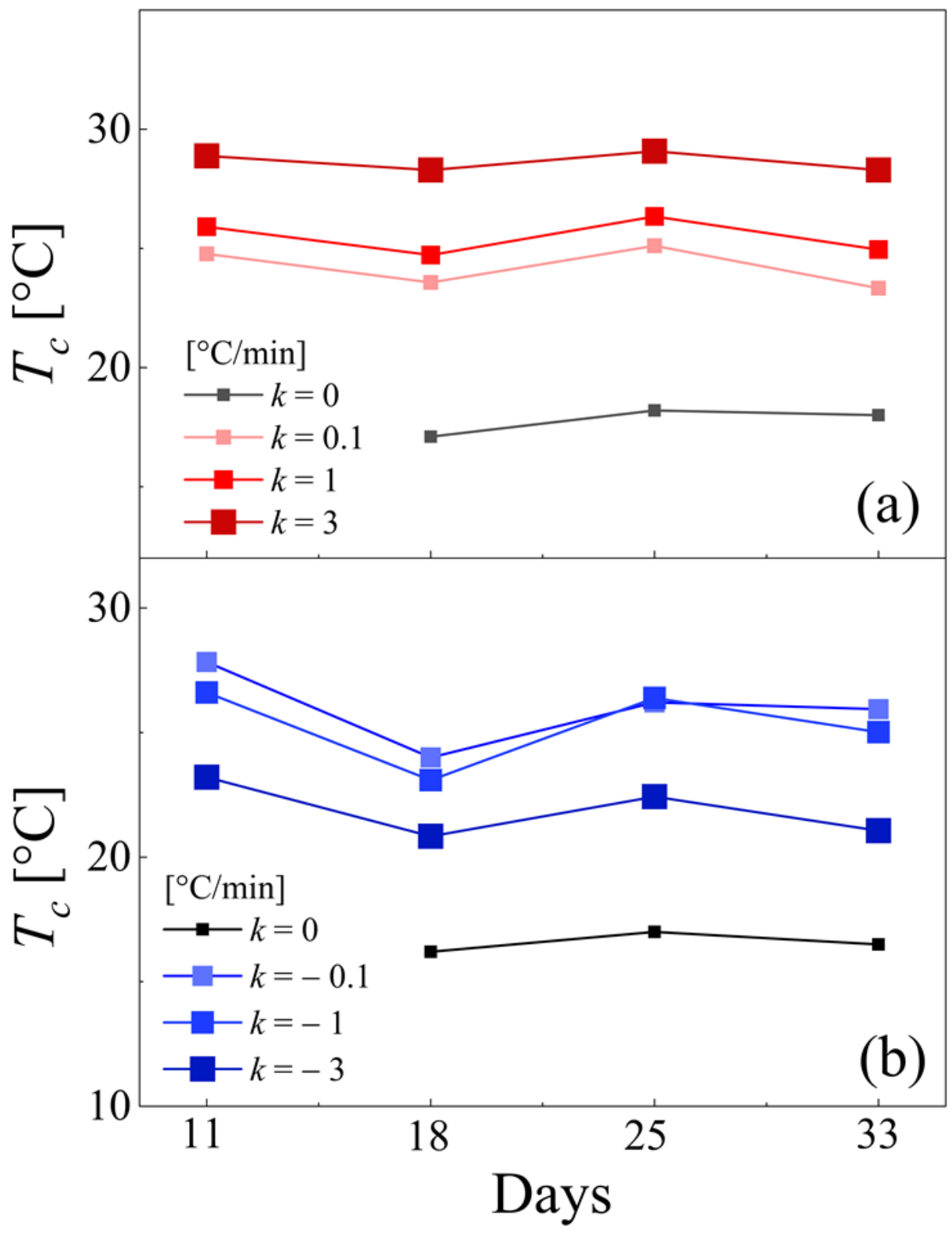

**Figure 8.** The variation in crossover temperature ($T_c$) is plotted as a function of the experimental day, illustrating the effect of both $k$ and experimental variability over time on the gelation behavior of 17 wt. % PF127 dispersion **(a)** during heating and **(b)** cooling cycles. Each solid square symbol represents $T_c$ at a particular $k$, where the increase in symbol size and change in colour represents an increase in $k$ of both heating and cooling cycles. The solid lines are added only to guide the eye.

Next, we investigate the effect of number of days since the date of preparation of 17 wt. % PF127 dispersion, keeping in mind that in industries such materials are routinely subjected to extended number of days for storage, and handling. To capture this essentially relevant aspect, we examine samples stored under refrigerated conditions at defined intervals of 11, 18, 25, and 33 days, with measurements performed after a regular interval (~7 days) to assess how time dependent changes may influence material response during thermal cycles. All rheological experiments involving different thermal ramp rates were conducted on a single day, which did not allow for multiple replicates under identical conditions within that timeframe and consequently, error bars are not reported for these measurements. However, to verify the consistency of the observed trends, the same set of experiments was repeated periodically over the course of one month. These repeated measurements consistently reproduced the qualitative behavior of the system, confirming that the trends are not artifacts of a single-day measurement and accurately reflect the inherent response of the material to changes in thermal ramp rate. Figure 8 a and b display the variation of $T_c$ as a function of experimental day for different $k$ during the heating and cooling cycle respectively. It is observed that for $k \neq 0$, $T_c$ across all ramp rates exhibits marginal variations for the explored number of days (day 11 to 33 after sample preparation), indicating minor sensitivity to storage period under finite ramp conditions. In contrast, $T_c$ at $k = 0$ of PF127 obtained from isothermal frequency sweep experiments remained essentially unchanged across all number of days explored in our study. This result further supports our inference that at $k = 0$ represents the intrinsic $T_c$ of the system, as the absence of an imposed finite ramps ($k \neq 0$) allows the system, to evolve near quasi-equilibrium conditions. By systematically evaluating $T_c$ over multiple days, our work addresses an overlooked but practically important question. It suggests that $T_c$ remains largely unchanged over storage time at all heating ramps, is valuable not only for ensuring reproducibility in laboratory studies but also for industrial applications where PF127 formulations may be prepared in advance, transported, or held prior to use.

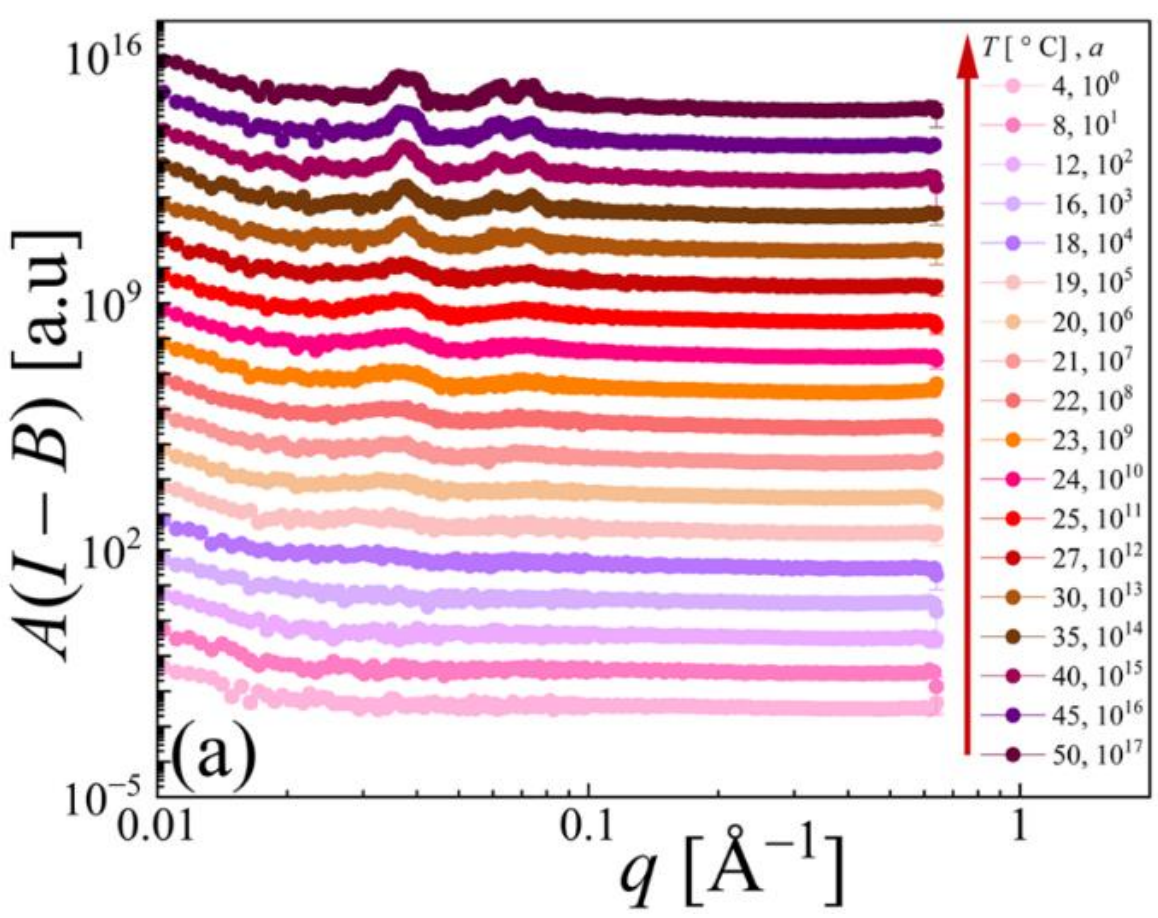

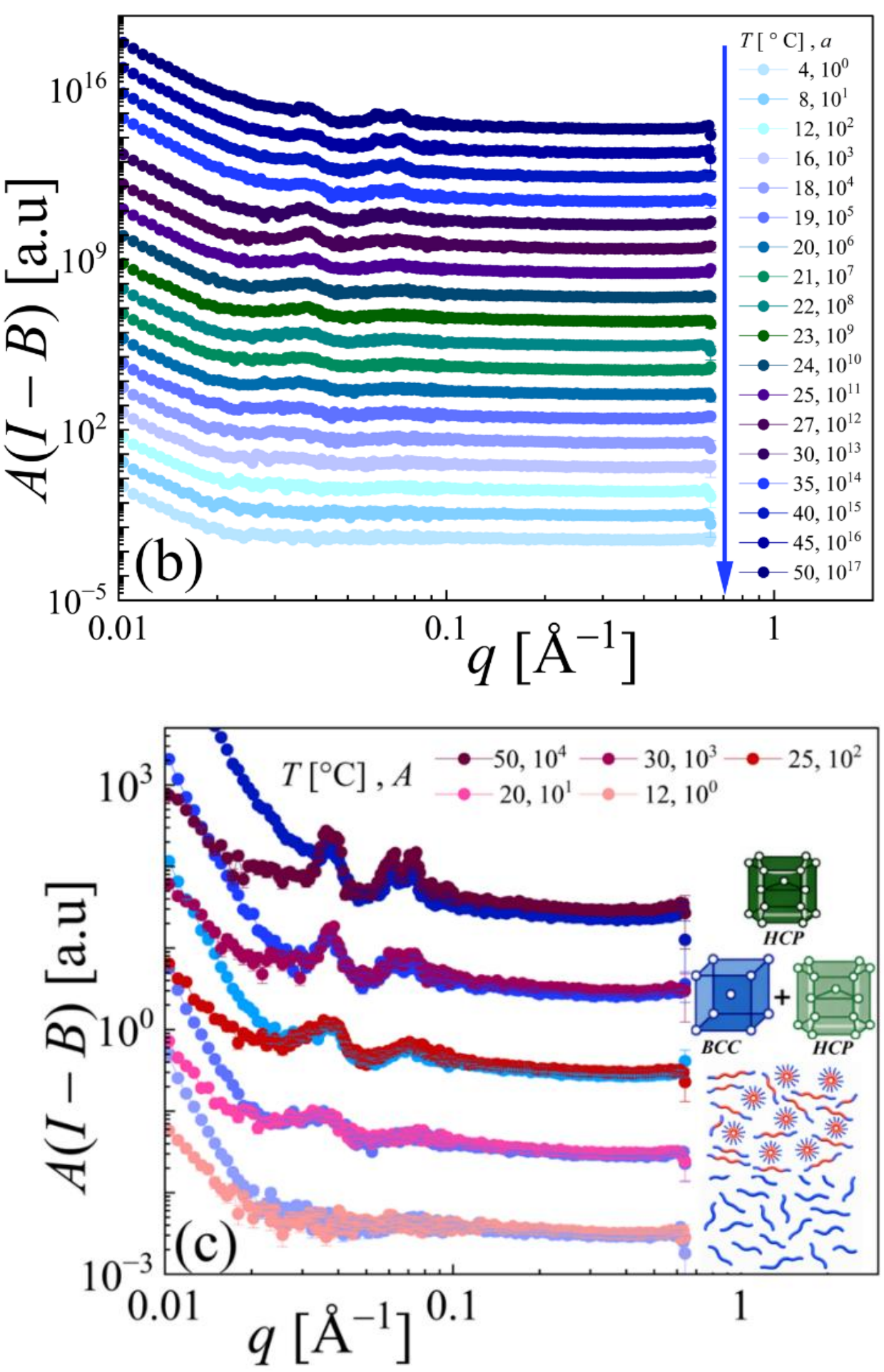

**Figure 9.** Small Angle X-ray Scattering (SAXS) intensity ($I - B$) is plotted as a function of scattering vector ($q$) for a 17 wt.% PF127 dispersion measured during increasing and decreasing temperature cycles from 4 °C to 50 °C. The scattering profiles are shown for **(a)** heating and **(b)** cooling cycles, and the curves are vertically shifted by a shift factor ($A$) for clarity. **(c)** discrete temperatures from heating (red curves) and cooling cycles (blue curves) at 12, 20, 25, 30, and 50 °C after applying the same vertical shift factor ($A$) for identical temperatures. The adjacent schematic shows microstructural evolution from unimers to micelles to ordered structures corresponding to the scattering profiles.

To understand the microstructural evolution of the 17wt.% PF127 system, we perform temperature dependent SAXS measurements during heating and cooling cycles, and the resulting scattering patterns are plotted in Figure 9 a and b respectively. During the SAXS measurements, the scattering data was acquired at a particular temperature after a thermal equilibration period of 10 minutes, ensuring thermal stability before data collection.

Scattering measurements under continuous temperature ramp, like in rheological experiments, could not be performed due to instrumental limitations associated with the SAXS setup. All the SAXS data represented in Figure 9 are vertically shifted by a shift factor (*A*). In Figure 9 a, which corresponds to the heating cycle, the scattering profiles remain indistinctive with no peaks from 4 °C to 18 °C, indicating a disordered micellar state. (49) The first peak begins to arise at $q = 0.035$ Å$^{-1}$ ($q_0$) at 20 °C, and as the temperature increases, a second peak appears at a higher $q$ value relative to $q_0$ at 23 °C. Subsequently, as the system reaches 30 °C, this second peak deconvolutes into two distinct peaks, resulting in a total of three well defined peaks. The intensities of three peaks increase steadily while their positions remain nearly unchanged, with further heating from 30 °C to 50 °C, suggesting progressive structural ordering.

To further understand the thermoreversibility of the PF127 system and to complement our rheological observations, we perform SAXS measurements during the cooling cycle following heating which is represented in Figure 9 b. To the best of our knowledge, scattering profiles acquired during the cooling step have not been previously reported for this system. Furthermore, these measurements were performed over a finely resolved temperature interval, enabling a more detailed and systematic characterization of the structural evolution. In Figure 9 b, the cooling cycle shows the reverse evolution, with the three distinct peaks weakening on cooling and gradually disappearing as the temperature decreases.

To enable a direct comparison of the scattering profiles obtained during the heating and cooling protocols, we construct a comparative plot at selected discrete temperatures. Figure 9 c represents SAXS profiles at five discrete temperatures (12, 20, 25, 30, and 50 °C), where data from both heating and cooling cycles at identical temperatures are overlapped by applying same *A* to the vertical axis of the curves, to improve clarity and enable direct comparison of peak positions. A schematic of the microstructural evolution from unimers to micelles to ordered structures at different temperatures has also been incorporated adjacent to the SAXS profiles in Figure 9 c. This overall representation gives an idea of the systematic structural progression that is consistently reproduced in both cycles, with no distinct peak at 12 °C, the emergence of a single peak at 20 °C, the appearance of second peak at 23 °C, the splitting of the second peak into two peaks at 30 °C, and lastly well-defined three peaks at 50

°C. Interestingly, we also observe that the peak intensity in the low $q$ region during heating cycle is slightly lower than those during cooling cycle suggesting larger particles/domains in solution during the cooling step. The close correspondence of peak positions and their evolution across heating and cooling indicates that the underlying structural organization is largely reversible, with minimal hysteresis in the length scales governing the assembly process, suggesting an overall thermoreversible behavior of the system. This observation further emphasizes that probing the structural organization under isothermal conditions provides a more reliable and representative approach for studying thermoreversible systems than measurements performed under continuously varying temperature. Such conditions allow the system sufficient time to reach equilibrium, thereby capturing the intrinsic structural features without the influence of kinetic constraints. In contrast, rate-dependent measurements, such as those conducted under continuous temperature ramps, are inherently affected by thermal history and kinetic limitations, and therefore may not fully resolve intermediate or equilibrium states.

From the obtained scattering curves, we use the relative peak positions to probe the ordered arrangement of the micelles, where features appearing in the mid $q$ region arise from interparticle correlations. The measured peak positions are analysed in terms of normalized ratios, $\frac{q}{q_0}$, where $q_0$ denotes the first peak position, and are compared with characteristic signatures reported for cubic lattices. We observe well-defined Bragg peaks at $\frac{q}{q_0}$ = 1, √3, and √4 in 17wt. % PF127 within the higher temperature range from 30 °C to 50 °C during both heating and cooling cycles. While no peaks are observed up to 19 °C (49), however the first and second Bragg peaks appearing between the intermediate temperature window from 20 °C and 23 °C at $\frac{q}{q_0}$ ≈ 1 and 1.7 do not correspond to any characteristic ratios associated with known crystalline lattices in the literature. This suggests the absence of well-defined long-range order in this regime. We attribute this to the system undergoing a transition from a homogenous system of unimers and micelles to a network structure in this temperature range, which is also consistent with our rheological measurements. In this intermediate state, the coexistence of multiple structural entities and the lack of periodic packing likely give rise to the broadened features in the scattering profiles. At 30 °C, the second peak deconvolutes into shoulders at $\frac{q}{q_0} \approx$ 1.53 and 1.82 alongside 1, marking body centred cubic (BCC) and hexagonally close packed (HCP) coexistence (50) as illustrated in the adjacent schematic in Figure 9 c. By 35 °C, BCC starts to fade and the pattern shifts towards 1: √3: √4 ratios. Peaks sharpen further at 40 °C with basal positions (100, 110, 200) indicating a nearly ideal HCP from 45 °C to 50 °C as represented via the adjacent schematic of Figure 9 c. On cooling, the same triplet re-emerges reversibly at identical temperatures, validating thermoreversible nature of polymer system. For clarity, HCP reflections are denoted using the simplified three-index (hkl) notation rather than the full Miller-Bravais convention. The absence of characteristic peaks at $\frac{q}{q_0}$ = √2 $\approx 1.41$ and $\approx 1.15$ excludes both BCC and FCC symmetries. (50) Peak sharpening and intensity growth with temperature reflect improving long-range order as micellar coronas contract and correlations strengthen. To the best of our knowledge, this is the first report of HCP micellar ordering as the dominant, thermally reversible equilibrium phase in a single-component aqueous PF127 at higher temperatures. We hypothesize that the soft PEO corona can reorganize more flexibly within the HCP geometry, while chain to chain variations further stabilize this packing.

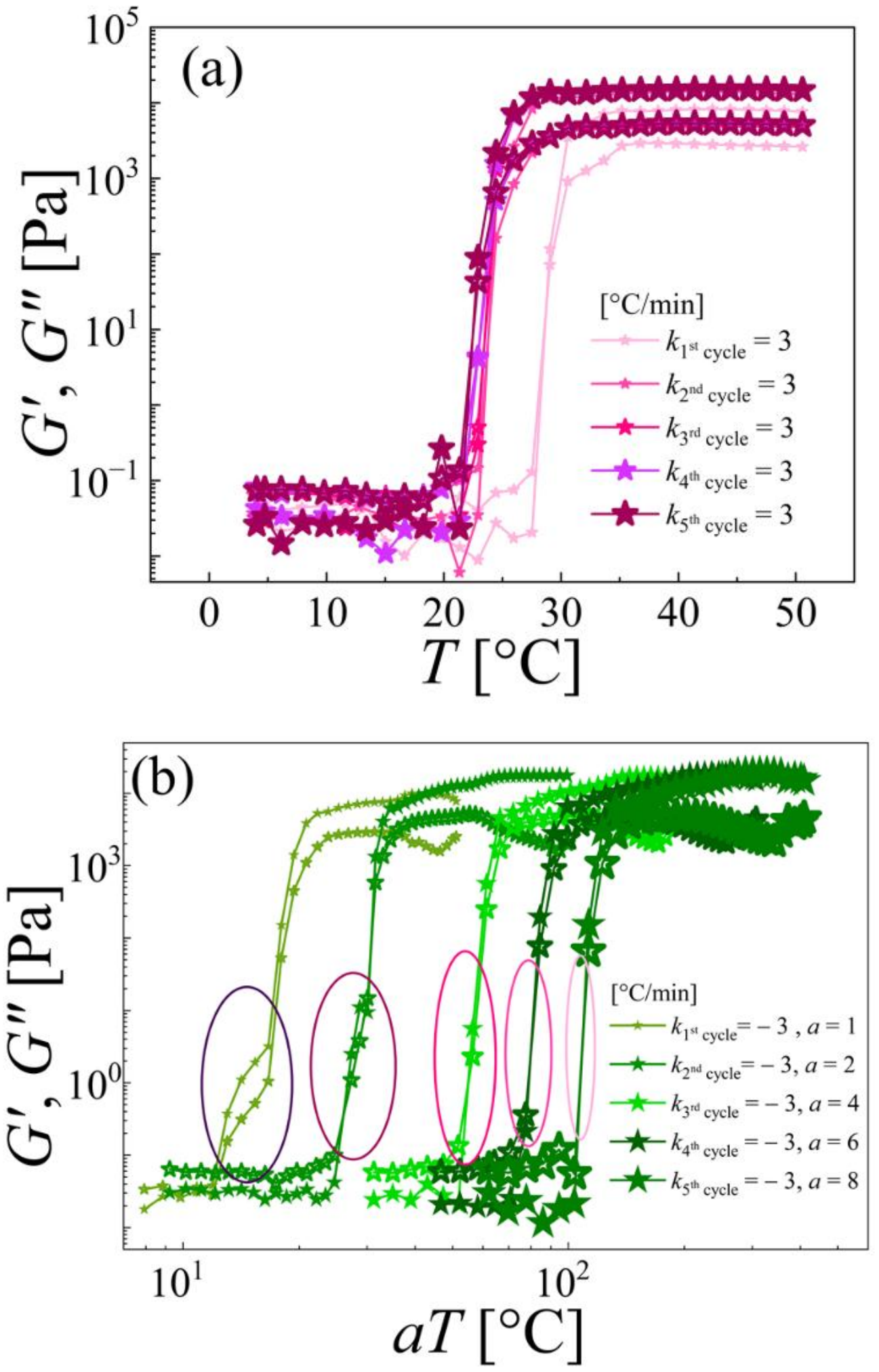

**Figure 10.** The variation in elastic modulus (*G′*, closed symbols) and viscous modulus (*G′′*, open symbols) of 17 wt. % PF127 dispersion is plotted as a function of temperature (*T*) at a constant ramp rate (*k*) of 3 °C/min across multiple **(a)** heating cycles and **(b)** cooling cycles, where the cooling curves are horizontally shifted by a shift factor (*a*) for clarity. The change in colour and increase in the size of the symbols correspond to successive heating and cooling cycles whereas the solid lines are included only to guide the eye.

We now examine the effect of thermal cycling on the viscoelastic response of the PF127 system, motivated by its direct application relevance, as such materials used in drug formulations are routinely subjected to repeated heating and cooling cycles. We perform experiments on PF127 dispersion at a fixed heating rate of $k$ = 3 °C/min to assess the effect of thermal cycling on the viscoelastic response where we examine the system over five consecutive thermal cycles performed sequentially. In Figure 10 a, we plot the variation in *G′*, and *G′′* as a function of temperature where each cycle is distinguished by progressively larger symbols, enabling clear visual identification of the thermal history. We observe that the first cycle exhibits distinct behavior, whereas subsequent cycles show reproducible and overlapping responses. This difference can be attributed to initial structural reorganization from a liquid sample. The sol-gel transition occurs at a higher temperature in the first cycle and subsequently, after the end of cooling cycle, the structure consists of more number of micelles which due to lack of any thermal equilibration at low temperature is unable to dissociate into individual unimers. Therefore, in the subsequent heating cycle, the transition occurs sooner and remains stabilized system in subsequent cycles. The overlapping data trends indicate minimal change in *G′*, and *G′′* trajectories as well as $T_c$ across all the repetitive cycles. The absence of significant drift in moduli values or transition temperatures across repeated cycles after the first cycle confirms that no cumulative hysteresis develops under the applied thermal cycling protocol. Therefore, the material maintains structural stability and demonstrates reliable, reversible gelation behavior under repetitive heating cycles. Such reversible response during successive heating cycles are consistent with prior rheological studies by Li *et al.*, on PF127 system stabilized using semi-interpenetrating network (sIPN). (32)

Furthermore, we studied the effect of multiple cooling cycles, to capture the evolution of the multiple step decay in *G′*, and *G′′* across repeated cooling cycles. Figure 10 b represents the results for five consecutive cooling cycles conducted at a constant $k = -3$ °C/min. Each successive cycle is marked by a progressively larger symbol, indicating increasing cycle number. To better visualize the overlapping trends and highlight the fading of the two-step decay, we have applied a horizontal shift factor '*a*' to the temperature axis to align the datasets and circular markers are used to emphasize the progressive disappearance of the two-step decay. It is evident from the plot that the distinct two-step decay, prominently observed during the first and second cooling cycles, progressively weakens with repeated cycling and completely disappears by the fifth cycle. In parallel, the initial sharp jumps in *G′* and *G″* curves become increasingly subdued, giving way to a smoother and more continuous decay of the moduli. The progressive weakening and eventual disappearance of the distinct two-step decay in *G′* and *G′′* with repeated cooling cycles likely reflects structural reorganization and partial stabilization of the polymer network during thermal cycling. During the first few cooling cycles, the system undergoes pronounced rearrangements as unimers, micelles, and transient network structures form and reorganize, giving rise to the characteristic two-step decay. However, repeated thermal cycling without any thermal equilibration at low temperatures does not allows the system to dissociate into individual unimers completely. The presence of the undissociated micelles and some clusters of micelles eliminates the formation of the metastable states that were responsible for the multi-step transition. Furthermore, some micelles may form more stable clusters during early cycles, diminishing the contribution of intermediate states in subsequent cycles.

From what we can gather, studies addressing the effect of repeated cooling cycles on thermoresponsive polymer systems remain limited. One of the few related reports by Li *et al.* (32) describes pronounced heating-cooling asymmetries in PF127 with semi-interpenetrating network system, where the sol to gel transition and its reverse occur over distinct temperature ranges on cooling versus heating. Interestingly, our measurements performed at a fixed cooling rate of − 3 °C/min reveal a distinct fading effect in successive cycles, where the characteristic multi-step decay in *G′* and *G″* becomes progressively less pronounced with repetitive cooling cycles. This indicates a progressive disruption of the usual microstructural reorganization or evolution without low-temperature rests across successive cooling cycles at same thermal ramp rate.

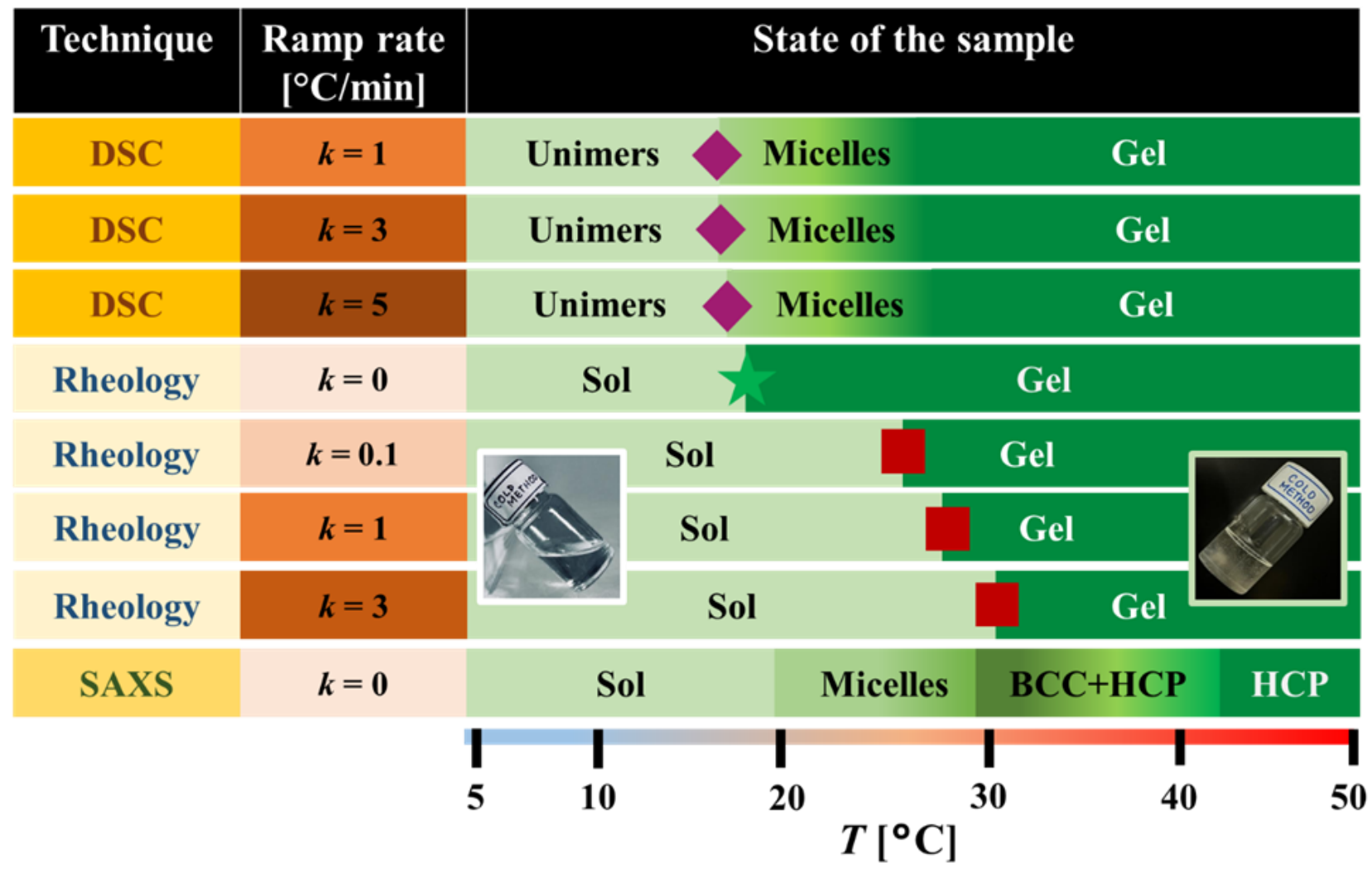

**Figure 11.** A scheme of the phase evolution of 17 wt.% PF127 determined through DSC, rheology, and SAXS has been represented. The system progresses from unimers to micelles

to a gel upon heating for all explored thermal ramp rates. At elevated temperatures, SAXS resolves the emergence of ordered micellar phases (BCC/HCP). The diamond, star and square symbols represent the micellization temperature $(T_m)$, critical gelation temperature $(T_g)$, and crossover temperature $(T_c)$ respectively.

To summarize, we present a phase diagram for 17 wt.% PF127 as characterized by the various experimental techniques explored in this work in Figure 11. The kinetic effect upon the transition temperature is clear from Figure 11 with measurable shifts arising from the applied thermal protocols. Furthermore, each experimental technique provides distinct and essential insights about the phase transition in the thermoresponsive polymer system. DSC experiments identify the onset of micellization while rheological experiments help in accurate identification of the sol-gel transition point. Moreover, SAXS at constant temperature renders insights about the microstructural arrangement of the micelles offering a structural perspective on the phase evolution. Altogether, these approaches establish a coherent and multidimensional understanding of the thermoresponsive behavior of PF127 solutions.

In summary, we present a comprehensive study of understanding the kinetic effect of thermal ramp rates and thermal equilibrium conditions on the phase behavior of thermoresponsive polymer solution. To begin with, consistent rate-dependent shifts in $T_m$ and peak intensity from DSC validates micellization in PF127 as an inherently kinetic process. Moreover, our rheological results highlight that the observed $T_c$ is not an intrinsic fixed parameter, but depends strongly on the applied thermal ramp rate, the heating or cooling cycle, sample age, and whether measurements are performed under thermal equilibrium conditions. These observations demonstrate the significant role of experimental protocol and kinetic history in governing the apparent phase transition behavior. We also present a simplified yet versatile mathematical framework capable of capturing the thermoresponsive viscoelastic behavior which can be extended to a broad class of materials. Connecting structure with macroscopic response, SAXS reveals a gradual transition from disordered unimers to HCP ordering that mirrors the rheological evolution of PF127. Overall, this work provides an important foundation for the careful identification of transition temperatures in thermoresponsive materials and underscores the necessity of accounting for kinetic effects when interpreting phase behavior and designing applications.

## Conclusions

This study presents a comprehensive investigation on rheological and structural characterization of aqueous PF127. It establishes that its thermoresponsive behavior is determined not only by concentration and temperature, but is significantly governed by thermal history and underlying kinetic pathways. Upon varying the thermal rate, the DSC measurements demonstrate systematic shift in both $T_m$ and peak intensity with heating and cooling rates.  These observations establish that micellization in PF127 is also a kinetically governed process. Our rheological investigations reveal a clear hysteresis between heating and cooling cycles. The sol to soft-solid transition during heating cycle occurs rapidly, whereas the cooling response follows a novel multi-step pathway involving the development of metastable intermediate states. The variation in thermal ramp rates during heating and cooling induces a systematic shift in $T_c$, confirming that it is a kinetically governed parameter influenced by micelle rearrangement, network formation timescales, and thermal equilibration. In contrast, $T_g$ obtained from isothermal frequency sweeps ($k$ = 0) represents a quasi-equilibrium transition and therefore, provides a more intrinsic and reliable measure of the true transition temperature compared to the rate-dependent $T_c$. Additionally, evaluation over multiple days shows that while $T_c$ may vary, the intrinsic $T_g$ under quasi-equilibrium conditions remains essentially invariant with explored number of days, across all cycles. Such observations are crucial for ensuring reproducibility in experimental studies and are equally relevant for industrial applications where PF127 formulations may be prepared in advance, stored, or transported prior to use. Furthermore, rheological measurements performed at a fixed cooling rate of − 3 °C/min reveal progressive fading of the characteristic multi step decay in $G'$and $G''$ with successive number of cycles. To describe these complex rheological responses in a quantitative manner, a modified kinetic model is applied to the rheological results and is validated. In addition, we assume that this modified kinetic model can be readily extended to describe multiple-step viscoelastic responses in other thermoresponsive polymers and polymer blends. Furthermore, SAXS analysis reveals a progressive evolution from disordered unimers at lower temperatures to HCP ordering at higher temperatures.

Collectively, these results show that the thermoresponsive and structural behavior of PF127 is strongly influenced by thermal history and temperature ramp rate, which are often overlooked in practical applications in industries. This integrated framework, combining calorimetry, rheological, scattering experimental results and kinetic modelling, provides a reliable base for understanding and designing thermoresponsive systems. It can be applied in areas such as drug delivery, injectable biomaterials, and bioprinting, where consistent and reproducible control over onset of gelation, structural reorganization, and thermal reversibility is critical for reliable product performance and manufacturing.

## Acknowledgements

We acknowledge the Centre for Soft and Biological Matter (CSBM) at IIT Madras for providing access to the SAXS facilities. We also thank the Thermal Analysis Laboratory (TAL), Department of Metallurgical and Materials Engineering, IIT Madras, for their support with the DSC experiments. We acknowledge the financial support from Anusandhan National Research Foundation (ANRF), Government of India (ANRF/ECRG/2024/001771/ENS), and HSBC grant (SB25261315CHHSBC009140) from the Center for Resource Efficiency, Recyclability and Circularity in Energy Transition, IIT Madras.

## Data Availability

The data that support the findings of this study are available from the corresponding author upon reasonable request.